\documentclass[pre,showpacs,showkeys,preprintnumbers,amsmath,amssymb,superscriptaddress,twocolumn]{revtex4}
\usepackage{graphicx}
\usepackage{amssymb,amsfonts,amsmath}

\def\ve{\varepsilon}
\def\pa{\partial\Omega}
\def\T{{\mathcal T}}
\def\D{{\mathcal D}}

\begin{document}

\title{Universal formula for the mean first passage time in planar domains}

\author{Denis~S.~Grebenkov}
 \email{denis.grebenkov@polytechnique.edu}

\affiliation{Laboratoire de Physique de la Mati\`{e}re Condens\'{e}e (UMR 7643), \\ 
CNRS -- Ecole Polytechnique, University Paris-Saclay, 91128 Palaiseau, France}

\date{Received: \today / Revised version: }

\begin{abstract}
We derive a general exact formula for the mean first passage time
(MFPT) from a fixed point inside a planar domain to an escape region
on its boundary.  The underlying mixed Dirichlet-Neumann boundary
value problem is conformally mapped onto the unit disk, solved
exactly, and mapped back.  The resulting formula for the MFPT is valid
for an arbitrary space-dependent diffusion coefficient, while the
leading logarithmic term is explicit, simple, and remarkably
universal.  In contrast to earlier works, we show that the natural
small parameter of the problem is the harmonic measure of the escape
region, not its perimeter.  The conventional scaling of the MFPT with
the area of the domain is altered when diffusing particles are
released near the escape region.  These findings change the current
view of escape problems and related chemical or biochemical kinetics
in complex, multiscale, porous or fractal domains, while the
fundamental relation to the harmonic measure opens new ways of
computing and interpreting MFPTs.
\end{abstract}



\maketitle 

How long does it take for diffusing species to exit from an irregular
domain or to initiate a reaction on a catalytic site or an enzyme?
Since the first contribution by Lord Rayleigh \cite{Rayleigh},
the first-passage phenomena have attracted much attention
\cite{Redner,Metzler,Benichou14,Holcman14}, and have been applied to
numerous chemical \cite{Hanggi90} and biological \cite{Bressloff13}
problems such as diffusion-influenced ligand binding to receptors on
cell surfaces \cite{Zwanzig91,Grigoriev02}, receptor trafficking in
synaptic membranes \cite{Holcman04}, diffusion in cellular
microdomains \cite{Schuss07}, or foraging strategies of animals
\cite{Viswanathan99}, to name but a few.  Most analytical results
were obtained for the mean first passage time (MFPT) to a small region
on the boundary, which can represent a specific target, a catalytic
germ, an active site, a channel or an exit to the outer space
\cite{Holcman14,Ward93,Singer06a,Singer06b,Singer06c,Pillay10,Cheviakov10,Cheviakov12,Caginalp12,Rupprecht15}.
For a ``regular'' planar domain $\Omega$ (whose perimeter
$|\pa|$ and linear size $|\Omega|^{\frac 12}$ are comparable, see
\cite{Holcman14}), the MFPT, averaged over uniformly distributed
starting points, was shown to be $\frac{|\Omega|}{\pi D} \bigl[\ln
(1/\ve) + O(1)\bigr]$, where $D$ is the diffusion coefficient,
$|\Omega|$ is the area of the domain, and $\ve = |\Gamma|/|\pa|\ll 1$
is the perimeter of the escape region $\Gamma$ divided by the
perimeter of the boundary $\pa$ \cite{Singer06a}.  In the special case
of a disk, Singer {\it et al.}  also showed that the MFPT from a fixed
starting point (e.g., the center) exhibits similar $\ln(1/\ve)$
behavior \cite{Singer06b}.  Since these seminal works, the logarithmic
divergence of the MFPT with respect to the normalized perimeter $\ve$
has become a common paradigm (see the review \cite{Holcman14} and
references therein).

In this letter, we show that this paradigm is incomplete for general
domains and can be strongly misleading when the starting point is
fixed.  We derive the exact formula for the MFPT from a fixed point
$x_0$ to a connected escape region $\Gamma$ on the boundary of any
simply connected (i.e., without ``holes'') planar domain $\Omega$.
This formula is not restricted to the narrow escape limit $\ve\ll 1$
and is valid for an arbitrary space-dependent diffusion coefficient.
Most importantly, we reveal an earlier unnoticed fundamental relation
between the MFPT and the harmonic measure of the escape region,
$\omega = \omega_{x_0}(\Gamma)$, i.e., the probability of arriving at
the escape region $\Gamma$ before hitting the remaining part of the
boundary \cite{Garnett}.

Before proceeding to rigorous results, we start with two examples of
``nonregular'' domains casting doubts on the normalized perimeter
$\ve$ as the universal small parameter.  If the domain is a thin long
rectangle $[0,L]\times [0,h]$, the MFPT to the left short edge from a
starting point $x_0 = (x_0^1,x_0^2)$ is equal to $(L x_0^1 - \frac12
[x_0^1]^2)/D$.  Being independent of $h$, this MFPT is thus not
determined by the normalized perimeter $\ve = \frac{h}{2(L+h)}$, even
if the latter is very small.  In the second example, one takes a disk
and replaces a small arc of its boundary by a very corrugated (e.g.,
fractal) curve.  Keeping the diameter $\delta$ of the modified escape
region small, one can make its perimeter arbitrarily large.  Since the
remaining part of the circle is fixed, the ratio $\ve =
|\Gamma|/|\pa|$ can be made close to $1$.  When $\delta$ is small, the
MFPT should be large, in spite the fact of $\ve \approx 1$.  These two
very basic examples illustrate the failure of the normalized perimeter
of the escape region as a determinant of the MFPT when the starting
point is fixed.  We will show that the natural characteristic that
substitutes the normalized perimeter $\ve$ is the harmonic measure
$\omega_{x_0}(\Gamma)$.

For Brownian motion starting from an interior point $x_0$ of a simply
connected planar domain $\Omega$, the MFPT $\T(x_0)$ to a connected
escape region $\Gamma$ on the boundary $\pa$ satisfies the backward
Fokker-Planck equation \cite{Gardiner}
\begin{equation}
\label{eq:FP_Omega}
\Delta \T(x_0) = - \frac{1}{D(x_0)}  \quad x_0\in\Omega,
\end{equation}
with mixed Dirichlet-Neumann boundary conditions
\begin{equation}
\label{eq:BC_Omega}
\T(x_0) = 0 \quad x_0\in \Gamma, \qquad \partial_n \T(x_0) = 0 \quad x_0\in \pa\backslash \Gamma ,
\end{equation}
where $\partial_n$ is the normal derivative, $\Delta$ is the Laplace
operator, and $D(x_0)$ is the space-dependent diffusion coefficient.
According to the Riemann mapping theorem, the unit disk $\D$ can be
mapped onto $\Omega$ by a conformal mapping
$\phi_{x_0}(z)~:~\D\to\Omega$.  We fix two parameters of the conformal
map by imposing that the origin of $\D$ is mapped onto the starting
point $x_0$: $\phi_{x_0}(0) = x_0$.  Since the conformal mapping
preserves the harmonic measure, the preimage of $\Gamma$ is an arc
$\gamma$ of the unit circle of length $2\pi \omega$.  Note that the
harmonic measure is fully determined by the conformal map.  The third
parameter of the conformal mapping is fixed by rotating the arc
$\gamma$ to be $(-\pi \omega,\pi \omega)$.  Setting $\tau(z) =
\T(\phi_{x_0}(z))$ for $z\in \D$, Eqs. (\ref{eq:FP_Omega},
\ref{eq:BC_Omega}) are transformed into
\begin{equation}
\left\{  \begin{array}{r l l}
 \Delta \tau(z) &= - |\phi'_{x_0}(z)|^2/D(\phi_{x_0}(z)) &  z\in \D, \\
 \tau(z) &= 0 & z \in \gamma,  \\ 
\partial_n \tau(z) &= 0 & z\in \partial \D\backslash \gamma.  \\  \end{array} \right.
\end{equation}
The solution of this mixed boundary value problem can be reduced to
dual trigonometric equations whose solutions are well documented
\cite{Sneddon}.  Skipping mathematical details (see SM1), we obtain
for any interior starting point $x_0 \in \Omega$
\begin{equation}
\label{eq:Tx0_D}
\T(x_0) =  \int\limits_\D  \frac{dz\, |\phi'_{x_0}(z)|^2}{D(\phi_{x_0}(z))} \biggl(- \frac{\ln |z|}{2\pi} + W_\omega(z) \biggr) \,  ,
\end{equation}
with
\begin{equation}
W_\omega(z) = \frac{1}{\pi} \ln \left(\frac{|1-z + \sqrt{(1-ze^{i\pi \omega})(1-ze^{-i\pi\omega})}|}{2\sin(\pi\omega/2)}\right),
\end{equation}
in which the most challenging ``ingredient'' of the problem, the mixed
boundary condition, is fully incorporated through the explicit
function $W_\omega(z)$.  The function $W_\omega(z)$ is universal; its
dependence on $\Omega$, $\Gamma$ and $x_0$ enters uniquely through the
harmonic measure $\omega = \omega_{x_0}(\Gamma)$.  To return to the
domain $\Omega$, the integration variable $z$ is changed to $x =
\phi_{x_0}(z)$, which yields
\begin{equation}
\label{eq:Tfinal}
\T(x_0) = \int\limits_\Omega  \frac{dx}{D(x)} \underbrace{\biggl(- \frac{\ln |\phi_{x_0}^{-1}(x)|}{2\pi} + 
W_\omega(\phi_{x_0}^{-1}(x)) \biggr)}_{\textrm{Green's function}}  .
\end{equation}

The exact solution (\ref{eq:Tfinal}) is our main result.  The two
terms are, respectively, (i) the MFPT from $x_0$ to the whole boundary
$\pa$, with $\frac{1}{2\pi} \ln |\phi^{-1}_{x_0}(x)|$ being the
Dirichlet Green's function in $\Omega$, and (ii) the contribution from
eventual reflections on the remaining part of the boundary,
$\pa\backslash \Gamma$, until reaching the escape region $\Gamma$.
The integral form of the solution $\T(x_0)$, which is valid for an
arbitrary function $1/D(x)$, allows one to interpret the expression in
parentheses in Eq. (\ref{eq:Tfinal}) as the Green's function of the
Laplace operator $-\Delta$ subject to mixed Dirichlet-Neumann boundary
condition (\ref{eq:BC_Omega}).  Numerical implementation of the exact
solution (\ref{eq:Tfinal}), its accuracy, and a comparison to
conventional numerical methods for computing MFPTs are discussed in
SM2.  While conformal mappings have been intensively used to solve
diffusion-reaction problems (e.g., see
\cite{Holcman14,Brady93,Koplik94,Koplik95,Hastings98,Davidovitch99,Blyth03,Bazant05,Chen11,Holcman13,Bazant16}
and references therein), this powerful technique is applied to
the mixed boundary value problem (\ref{eq:FP_Omega},
\ref{eq:BC_Omega}) for the first time.  Prior to this work, no exact
solution of this MFPT problem was available, except for a few simple
domains \cite{Singer06b,Caginalp12,Rupprecht15}.  While the proposed
approach is limited to planar Brownian diffusion (see the further
discussion in SM1), the universality of the exact solution
(\ref{eq:Tfinal}) results from the existence of a conformal map for
any simply connected planar domain, even with a very irregular (e.g.,
fractal) boundary.  The only mathematical restriction on the domain is
that the original problem (\ref{eq:FP_Omega}, \ref{eq:BC_Omega})
should be well defined.  In the case of the unit disk with a constant
diffusivity, our general formula (\ref{eq:Tfinal}) reduces to the
earlier result \cite{Caginalp12}
\begin{equation}
\T(x_0) = \frac{1 - |x_0|^2}{4D} + \frac{\pi}{D} W_{\ve}(x_0)
\end{equation}
(see SM3 for the derivation). 

The general solution (\ref{eq:Tfinal}) allows one to investigate, for
the first time, the impact of heterogeneous diffusivity in MFPT
problems defined in nontrivial domains.  In spite of its evident
importance for biological systems in which the spatial heterogeneity
is ubiquitous, only a few analytical results about one-dimensional
motion with space-dependent diffusivity are available
\cite{Cherstvy13}.  Even for simple domains such as disks or
rectangles, the spatial dependence of the diffusion coefficient can
prohibit the separation of variables or, in the most favorable cases,
lead to complicated differential equations whose solutions cannot be
expressed in terms of usual special functions.  In this light, it is
remarkable that the formula (\ref{eq:Tfinal}) provides an exact
solution even for the space-dependent diffusivity $D(x)$.  One can
investigate, e.g., how the lower diffusivity in the actin cortex near
the plasma membrane would affect the overall MFPT in flat living cells
adhered to a surface, as well as the effect of corrals on membrane
diffusion.

When the harmonic measure of the escape region is small, one can
derive the perturbative expansion of the MFPT (see SM4):
\begin{equation}
\label{eq:TfinalA}
\begin{split}
\T(x_0) & = \frac{|\Omega|}{\pi D_h} \ln (1/\omega) \\
& + \biggl(\frac{|\Omega| \ln(2/\pi)}{\pi D_h} + \frac{1}{2\pi} 
\int\limits_\Omega \frac{dx}{D(x)} \ln \frac{|1-\phi^{-1}_{x_0}(x)|^2}{|\phi^{-1}_{x_0}(x)|} \biggr) \\
& + \omega^2 \biggl(\frac{\pi |\Omega|}{24 D_h} - \int\limits_\Omega dx \frac{V_2(\phi^{-1}_{x_0}(x))}{D(x)} \biggr) + O(\omega^4) , \\
\end{split}
\end{equation}
with 
\begin{equation}
V_2(z) = - \frac{\pi}{8} \biggl(\frac{z}{(1-z)^2} + \frac{\bar{z}}{(1-\bar{z})^2}\biggr) .   
\end{equation}
The first logarithmic term appears as the leading contribution that
involves the harmonic measure of the escape region, $\omega =
\omega_{x_0}(\Gamma)$, the area of the domain, $|\Omega|$, and the
harmonic mean of the space-dependent diffusion coefficient:
\begin{equation}
\label{eq:Dh}
\frac{1}{D_h} = \frac{1}{|\Omega|} \int\limits_\Omega  \frac{dx}{D(x)} \, 
\end{equation}
(if $D(x) = D$ is constant, then $D_h = D$).  The expansion
(\ref{eq:TfinalA}) and its leading term $\frac{|\Omega|}{\pi D_h}
\ln(1/\omega_{x_0})$ present the second main result that allows one to
estimate the MFPT directly through the harmonic measure.

\begin{figure}
\begin{center}
\includegraphics[width=85mm]{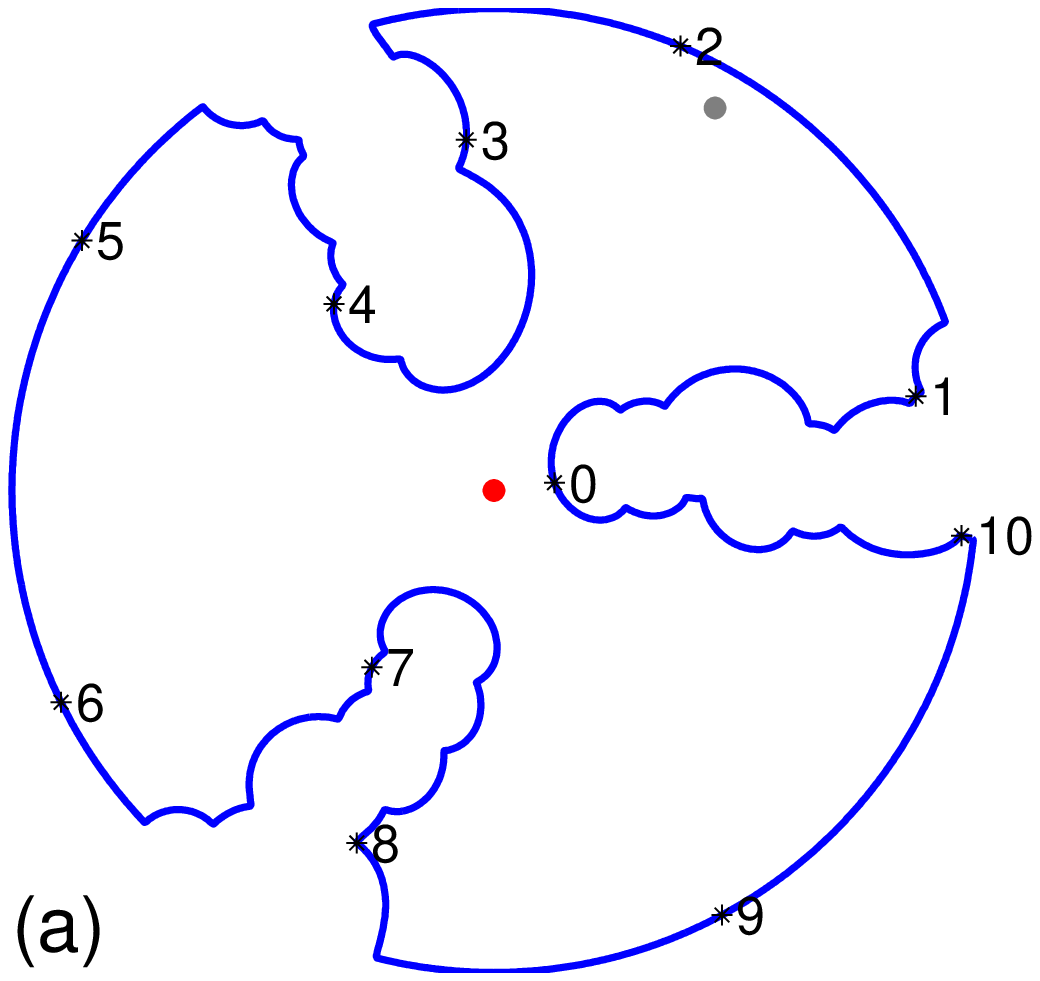}
\includegraphics[width=85mm]{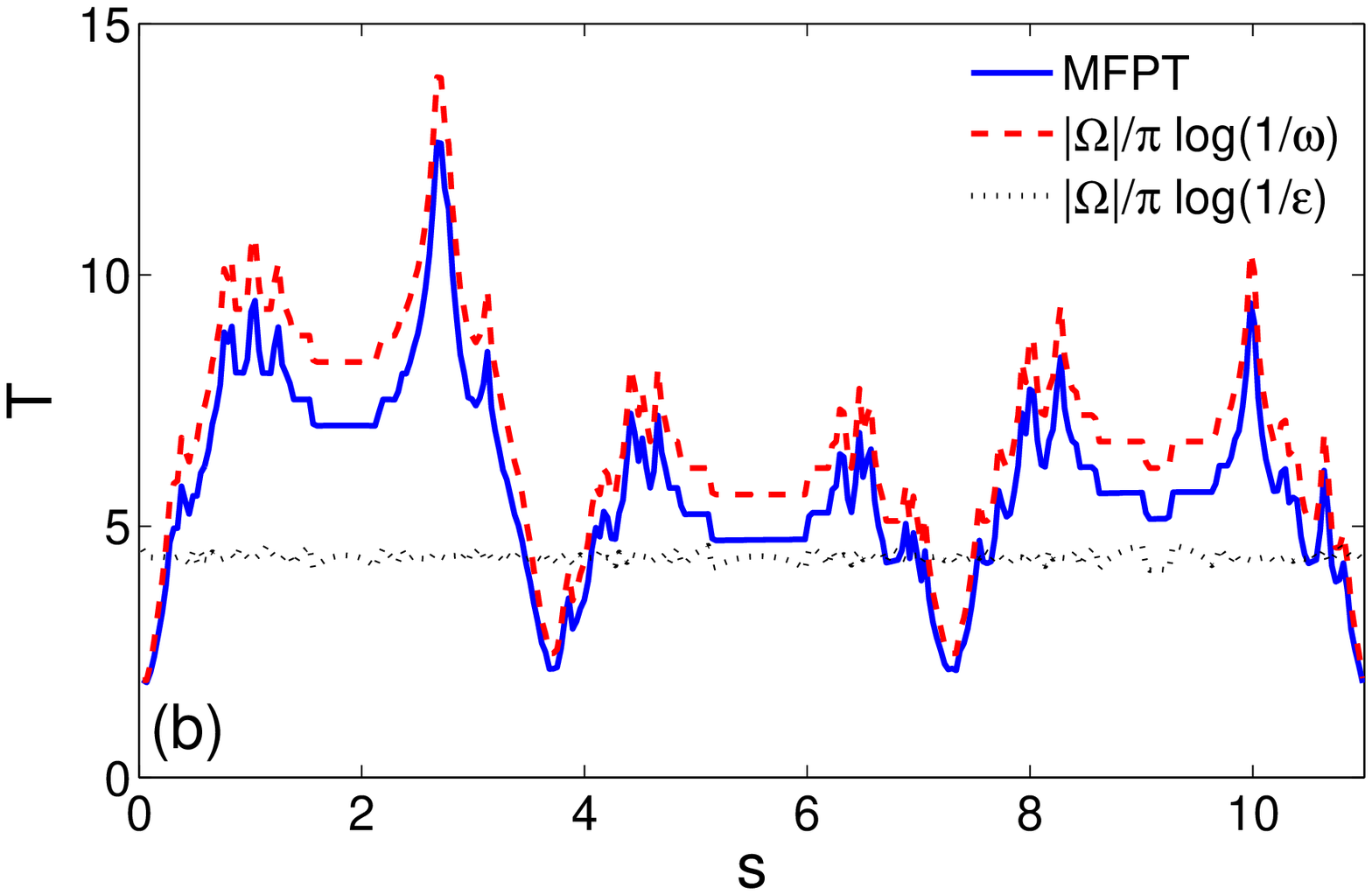}
\includegraphics[width=85mm]{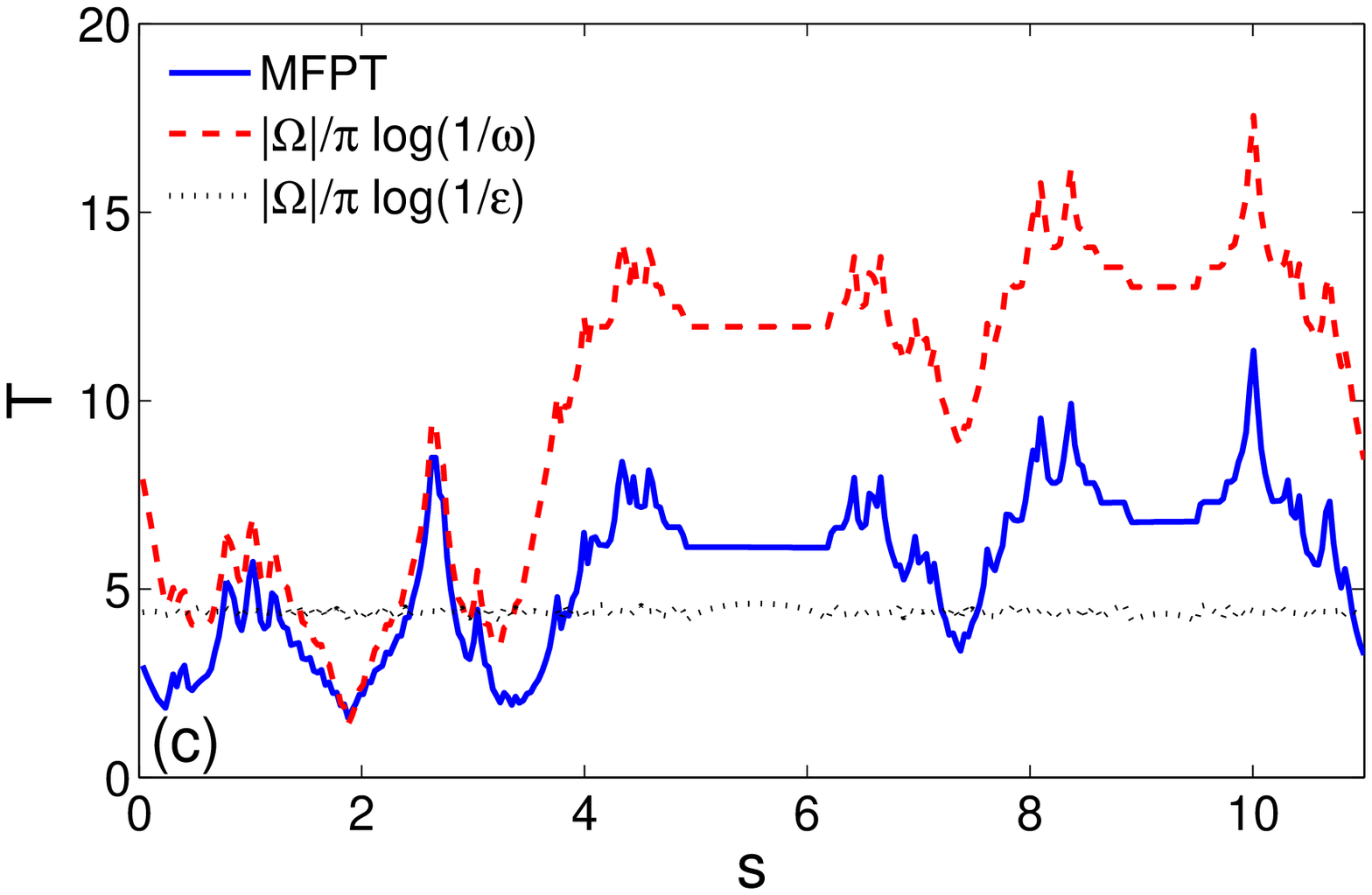}
\end{center}
\caption{
(Color online) {\bf (a)} An irregular domain obtained by iterative
conformal maps (see SM5).  The red and gray circles indicate two
considered starting points: $x_0 = 0$ and $x_0 = 0.4590 + 0.7936i$;
the asterisks with indicated curvilinear coordinates present
``milestones'' along the boundary; the perimeter, the diameter, and
the area of the domain are $10.96$, $2$, and $2.39$, respectively.
{\bf (b,c)} The MFPT $\T(x_0)$ (solid line), the leading term
$\frac{|\Omega|}{\pi D} \ln(1/\omega_{x_0}(\Gamma_s))$ (dashed line),
and the conventional leading term $\frac{|\Omega|}{\pi D}\ln(1/\ve)$
(dotted line) as functions of the location $s$ of the escape region
$\Gamma_s$ on the boundary of the shown domain, with $x_0 = 0$ {\bf
(b)} and $x_0 = 0.4590 + 0.7936i$ {\bf (c)}.  Note that small
fluctuations of the dotted line are related to small variations in the
perimeter of the escape region due to discretization.  Here, we set $D
= 1$ and use dimensionless units. }
\label{fig:MFPT}
\end{figure}

In fact, when the starting point $x_0$ is not too close to the
boundary, the leading term provides a good approximation to the MFPT.
To illustrate this point, we generated a planar domain in
Fig. \ref{fig:MFPT}a by iterative conformal maps (see SM5).  In order
to emphasize that the normalized perimeter $\ve$ of the escape region
must be replaced by the harmonic measure, we fix $\ve = 0.035$ and
then move the escape region $\Gamma_s$ of perimeter $\ve |\pa|$ along
the boundary of the domain, where $s$ is the curvilinear coordinate of
the center of $\Gamma_s$.  Figure \ref{fig:MFPT}b shows $\T(x_0)$ from
Eq. (\ref{eq:Tfinal}) and its leading term $-\frac{|\Omega|}{\pi D}
\ln \omega_{x_0}(\Gamma_s)$ from Eq. (\ref{eq:TfinalA}), as functions
of the location $s$ of the escape region $\Gamma_s$ on the boundary
for $x_0 = 0$ (we set $D = 1$ and use dimensionless units).  As
expected, the MFPT significantly varies when the escape region moves,
showing the dependence on the distance between $x_0$ and $\Gamma$ and
on the shape of the domain.  For instance, the corner at the
curvilinear location $s \approx 2.7$ is difficult to access so that
the related harmonic measure is very small, while the MFPT exhibits a
prominent peak rising up to $14$.  In turn, three escape regions at $s
\approx 0,\, 3.7,\, 7.3$ are the closest to the starting point and
thus easily accessible, resulting in the minima of the MFPT.  One can
see that the leading term with the harmonic measure provides an
excellent approximation to the MFPT.  In turn, the conventional
leading term $\frac{|\Omega|}{\pi D}\ln (1/\ve)$ (which is independent
of the location) is a poor estimate, in spite of the fact that $\ve$
is small.  We conclude that when the starting point is fixed, the
harmonic measure of the escape region should substitute the normalized
perimeter as the natural small parameter, at least in two dimensions.

When the starting point $x_0$ is close to the boundary, the
logarithmic term overestimates the MFPT, as illustrated in
Fig. \ref{fig:MFPT}c for the starting point $x_0 = 0.4590 + 0.7936i$.
One can see that the logarithmic term accurately captures the behavior
of the MFPT when the escape region is close to the starting point ($s$
between 1 and 3) while a significant but nearly constant deviation
appears for distant escape regions.  Here, the contribution from the
next-order terms in Eq. (\ref{eq:TfinalA}), in particular, that of
order $O(1)$, is comparable to the logarithmic term.

This situation can also be illustrated on the example of a thin
rectangle $[0,L]\times [0,h]$ that we discussed at the beginning.  In
fact, the harmonic measure $\omega$ of the short edge on the left,
seen from a point $(x_0^1,h/2)$, is approximately $\exp(- \pi
x_0^1/h)$ (see SM3).  Substitution of this expression into the first
term of Eq. (\ref{eq:TfinalA}) yields $Lx_0^1/D$, which is close to
the exact MFPT $(Lx_0^1 - \frac12 [x_0^1]^2)/D$.  This simple example
illustrates that (i) even if the harmonic measure $\omega$ is very
small, the contribution from the remaining terms in
Eq. (\ref{eq:TfinalA}) can still be significant (e.g., the term
$-\frac12 [x_0^1]^2/D$ in this example), and (ii) this MFPT does not
scale with the area of the domain.  Interestingly, even though the
area of the rectangle, $|\Omega| = Lh$, stands in front of the
logarithmic term in Eq. (\ref{eq:TfinalA}), the thickness $h$ is then
removed by the factor $1/h$ coming from $\ln \omega$.  The last
observation challenges another common paradigm that the MFPT is
proportional to the area of the domain.  In particular, if particles
are released from a fixed point near the escape region, most of them
find it very rapidly.  Would the contribution to the MFPT from a few
particles that miss the escape region at the beginning and thus
explore the whole domain, ensure the scaling $\T(x_0)\propto
|\Omega|$?  The answer is in general negative as discussed in SM6 and
illustrated above in the case of a thin rectangle.

We can now revise the second example mentioned above, namely, a disk
with a small but highly corrugated arc.  According to Makarov's
theorem, the information dimension of the harmonic measure is equal to
$1$ for planar connected sets \cite{Makarov85,Jones88}, implying that
$\omega_{x_0}(\Gamma)$ scales with the diameter of $\Gamma$
\cite{Sapoval94}.  In other words, the harmonic measure of the
corrugated arc, seen from a distant point $x_0$, is determined by the
diameter $\delta$ of the arc, not the perimeter $\ve$.  We thus
recover the intuitively expected behavior $\frac{|\Omega|}{\pi D} \ln
(1/\delta)$ when $\delta$ is small.  In turn, the conventional formula
$\frac{|\Omega|}{\pi D} \ln (1/\ve)$ is again strongly misleading.  We
emphasize that $\omega$ and $\ve$ are in general unrelated; e.g., a
set may have an arbitrarily small harmonic measure and arbitrarily
large perimeter, and vice versa.

These examples illustrate generic features of the MFPT in porous media
with long channels or fjords, and in domains with irregular
boundaries, in contrast to earlier works that dealt with very regular
domains whose perimeter $|\pa|$ and linear size $|\Omega|^{\frac 12}$
were comparable (see \cite{Holcman14}).  In the latter case, the
harmonic measure of the escape region is proportional to its Lebesgue
measure (i.e., the perimeter), recovering the conventional $\ln
(1/\ve)$ behavior, the proportionality coefficient (the harmonic
measure density) and the related dependence on the starting point
$x_0$ being ``hidden'' in the $O(1)$ term.
We stress that even for such regular domains, the dependence on the
starting point can be strong and provide the dominant contribution to
the MFPT, as illustrated in Fig. \ref{fig:MFPT}b,c.  For instance, one
can think of two pores connected by a very narrow channel.  If the
particle starts inside one pore while the escape region is located on
the boundary of the other pore, the MFPT can be made arbitrarily large
by controlling the channel width, even if the escape region remains
large whereas the condition $|\pa|\sim |\Omega|^{\frac 12}$ is
satisfied.  Only the average of $\T(x_0)$ over the starting point
$x_0$ might recover the $\ln (1/\ve)$ dependence in the leading term.
However, in many applications, the source of particles and the escape
region are well separated (e.g., viruses entering the cell at the
membrane and searching for the nucleus, or molecules released near the
nucleus and searching to escape through the membrane).  The proposed
formula (\ref{eq:Tfinal}) thus yields a powerful tool to investigate
these search and escape phenomena.

The uncovered relation between the MFPT and the harmonic measure
brings new opportunities.  On one hand, one can profit from numerous
analytical and numerical results known for the harmonic measure on
irregular boundaries
\cite{Garnett,Makarov85,Jones88,Cates87,Meakin87,Mandelbrot90,Evertsz91,Evertsz92,Makarov99,Duplantier99,Grebenkov05,Grebenkov05c,Bettelheim05,Grebenkov15}.
In particular, the concept of diffusion screening
\cite{Meakin85,Sapoval94} that has found numerous implications for
heterogeneous catalysis \cite{Filoche08}, fluid flow in rough channels
\cite{Andrade07,Bazant16}, and transport phenomena in biological
systems \cite{Felici03,Grebenkov05b,Gill11,Maoy12} can now be applied
to the MFPT.  For instance, the harmonic measure of an escape region
at the bottom of a fjord can exhibit various types of decay with the
``depth,'' depending on the shape \cite{Garnett,Evertsz91}.  Similar
dependences are thus expected for the MFPT.  On the other hand, the
conformal mapping is a powerful analytical and numerical technique to
represent the geometric complexity of a domain through analytic
properties of the mapping function $\phi_{x_0}(z)$ \cite{Schinzinger}.
The unit disk can be conformally mapped onto any polygon either by a
Schwarz-Christoffel formula \cite{Driscoll,Banjai03} or by a
``zipper'' algorithm \cite{Marshall07}.  Once the conformal map is
constructed, finding the MFPT in complex domains is reduced to
computing the integrals in Eqs. (\ref{eq:Tx0_D}) or (\ref{eq:Tfinal}).
Most importantly, the proposed approach is not limited to regular
domains (with $|\pa|\sim |\Omega|^{\frac 12}$) and allows one to study
the MFPT for irregular (e.g., fractal) boundaries and branching
domains with long rough channels and large surface-to-volume ratios
that are relevant for most applications.  This approach opens thus a
new field of research on first passage times and related chemical or
biochemical kinetics in complex, multiscale, and porous media.

\begin{acknowledgments}
The author acknowledges support under Grant No. ANR-13-JSV5-0006-01 of
the French National Research Agency, and Dr. A. Rutenberg and
Dr. D. Belyaev for fruitful discussions.
\end{acknowledgments}


\pagebreak


\textbf{Supplementary Material for the Letter ``Universal formula for the mean first passage time in planar domains''}%
\label{SM}
\setcounter{equation}{0}
\setcounter{section}{0}
\setcounter{figure}{0}
\setcounter{table}{0}
\setcounter{page}{1}
\setcounter{footnote}{0}

\makeatletter
\renewcommand{\thepage}{S\arabic{page}}
\renewcommand{\theequation}{S\arabic{equation}}
\renewcommand{\thefigure}{S\arabic{figure}}
\renewcommand{\thesection}{SM\arabic{section}}


\section{Derivation of the main formula}

In this Section, we present the derivation of the exact formula for
the MFPT $\T(x_0)$.  

The function $\phi_{x_0}(z)$ that conformally maps the unit disk $\D$
onto the domain $\Omega$, transforms the original boundary value
problem for the MFPT $\T(x_0)$ in $\Omega$,
\begin{equation}
\label{eq:SFP_Omega}
\left\{ \begin{array}{r l l}
\Delta \T(x_0) & = - 1/D(x_0) &  x_0\in\Omega, \\
\T(x_0) & = 0  &  x_0\in \Gamma, \\ 
\partial_n \T(x_0) & = 0 & x_0\in \pa\backslash \Gamma , \\  \end{array} \right.
\end{equation}
into another boundary value problem for $\tau(z) = \T(\phi_{x_0}(z))$
in $\D$
\begin{equation}
\label{eq:FP_D}
\left\{  \begin{array}{r l l}
 \Delta \tau(z) &= f(z) &  z\in \D, \\
 \tau(z) &= 0 & z \in \gamma,  \\ 
\partial_n \tau(z) &= 0 & z\in \partial \D\backslash \gamma,  \\  \end{array} \right.
\end{equation}
where 
\begin{equation}
f(z) = - \frac{|\phi'_{x_0}(z)|^2}{D(\phi_{x_0}(z))} \,, 
\end{equation}
and the arc $\gamma = (-\pi \omega,\pi \omega)$ is the pre-image of
the escape region $\Gamma$.  Here $0 \leq \omega \leq 1$ is the
harmonic measure of the escape region seen from the starting point
$x_0$ \cite{Garnett}.  Since we imposed $\phi_{x_0}(0) = x_0$ for a
fixed $x_0$, we have $\T(x_0) = \tau(0)$ for this particular starting
point.  This choice of the conformal map greatly simplifies the
following derivation.

The solution of the problem (\ref{eq:FP_D}) can be represented as
$\tau(z) = u(z) + v(z)$, with $v(z)$ satisfying the inhomogeneous
Laplace equation with Dirichlet boundary condition,
\begin{equation}
\label{eq:v}
\left\{  \begin{array}{r l l}
\Delta v(z) & = f(z)  & z\in \D, \\
 v(z) & = 0 & z \in \partial \D, \\  \end{array} \right.
\end{equation}
while $u(z)$ satisfying homogeneous Laplace equation with mixed
Dirichlet-Neumann boundary condition:
\begin{equation}
\label{eq:u}
\left\{  \begin{array}{r l l}
\Delta u(z) & = 0  & z\in \D, \\
 u(z) & = 0 & z \in \gamma,  \\
\partial_n u(z) &= -\partial_n v(z) & z\in \partial \D\backslash \gamma .\\  \end{array} \right.
\end{equation}

The solution of the problem (\ref{eq:v}) is
\begin{equation}
\label{eq:vz}
v(z) = \int\limits_\D dz_0 \, G(z,z_0) \, f(z_0)\,, 
\end{equation}
where $G(z,z_0)$ is the Green's function on the unit disk,
\begin{equation}
G(z,z_0) = \frac{1}{2\pi} \ln \biggl| \frac{z-z_0}{1-\bar z_0  z}\biggr| , 
\end{equation}
and bar denotes complex conjugate.  In what follows, we focus on the
solution of the homogeneous Laplace equation (\ref{eq:u}).

In order to reduce this problem to dual trigonometric equations, we
split the solution in ``symmetric'' and ``antisymmetric'' parts, $u =
u_+ + u_-$, which satisfy
\begin{equation}
\left\{  \begin{array}{r l l}
\Delta u_\pm(z) &= 0 & z\in \D, \\
 u_\pm(z) &= 0  & z \in \gamma, \\
 \partial_n u_\pm(z) & = -\frac12 \bigl[\partial_n v(z) \pm \partial_n v(\bar{z})\bigr] &  z\in \partial \D\backslash \gamma. \\  \end{array} \right.
\end{equation}
In polar coordinates, $u_+(r,\theta)$ and $u_-(r,\theta)$ are
respectively symmetric and antisymmetric functions with respect to
reflection $\theta \to -\theta$.

We search the symmetric solution as
\begin{equation}
u_+(r,\theta) = \frac12 a_0 + \sum\limits_{n=1}^\infty a_n r^n \cos n \theta,
\end{equation}
with unknown coefficients $a_n$ fixed by boundary conditions:
\begin{equation}
\label{eq:dual}
\begin{split}
\bigl(\partial_r u_+(r,\theta)\bigr)_{r=1} = \sum\limits_{n=1}^\infty n a_n \cos n\theta &= F(\theta) \quad 0 \leq \theta \leq c ,  \\
u_+(1,\theta) = \frac12 a_0 + \sum\limits_{n=1}^\infty a_n \cos n\theta &= 0  \qquad c \leq \theta \leq \pi,  \\
\end{split}
\end{equation}
where 
\begin{equation}
\label{eq:c}
c = \pi - \pi \omega,
\end{equation}
\begin{equation}
\label{eq:F_Poisson}
\begin{split}
F(\theta) & = -\frac12 \bigl[\partial_r v(r,\theta) + \partial_r v(r,-\theta)\bigr]_{r=1} \\
& = - \frac12\int\limits_\D dz_0 \, \bigl[\omega_{z_0}(\theta) + \omega_{z_0}(-\theta)\bigr] \, f(z_0) \,,
\end{split}
\end{equation}
and
\begin{equation}
\label{eq:Poisson}
\omega_{z_0}(\theta) = [\partial_r G(z,z_0)]_{r=1} = \frac{1-r_0^2}{2\pi(1 - 2r_0 \cos(\theta-\theta_0) + r_0^2)}
\end{equation}
is the Poisson kernel (i.e., the harmonic measure density in the unit
disk), with $z_0 = r_0 e^{i\theta_0}$.  Note that we temporarily
rotated the arc $\gamma$ to be $(c,2\pi-c)$ that is equivalent to
replacing $\theta$ by $\pi -\theta$.  We will rotate it back at the
end of derivation.

The solution of the dual equations (\ref{eq:dual}) is given in
\cite{Sneddon} (see Eqs. (5.4.54,~5.4.55) with $G(\theta) = 0$ and
$\lambda = 0$):
\begin{equation}
\begin{split}
\frac{a_0}{2} & = \frac{\sqrt{2}}{\pi} \int\limits_0^c d\theta \frac{\sin(\theta/2)}{\sqrt{\cos \theta - \cos c}} \int\limits_0^\theta d\theta' ~ F(\theta') \\
& = - \int\limits_\D dz_0 \, W_\omega(z_0) \,  f(z_0) \,, \\
\end{split}
\end{equation}
with
\begin{equation}
\label{eq:Wdef}
\begin{split}
W_\omega(z_0) & = \frac{1}{\pi \sqrt{2}} \int\limits_0^c d\theta \frac{\sin(\theta/2)}{\sqrt{\cos \theta - \cos c}} \\
& \times \int\limits_0^\theta d\theta'  \bigl[\omega_{z_0}(\theta') + \omega_{z_0}(-\theta') \bigr] .\\
\end{split}
\end{equation}
One can also get the other coefficients $a_n$ but they are not needed
since we are only interested in $u_+(0,0)$.

\begin{figure}
\begin{center}
\includegraphics[width=85mm]{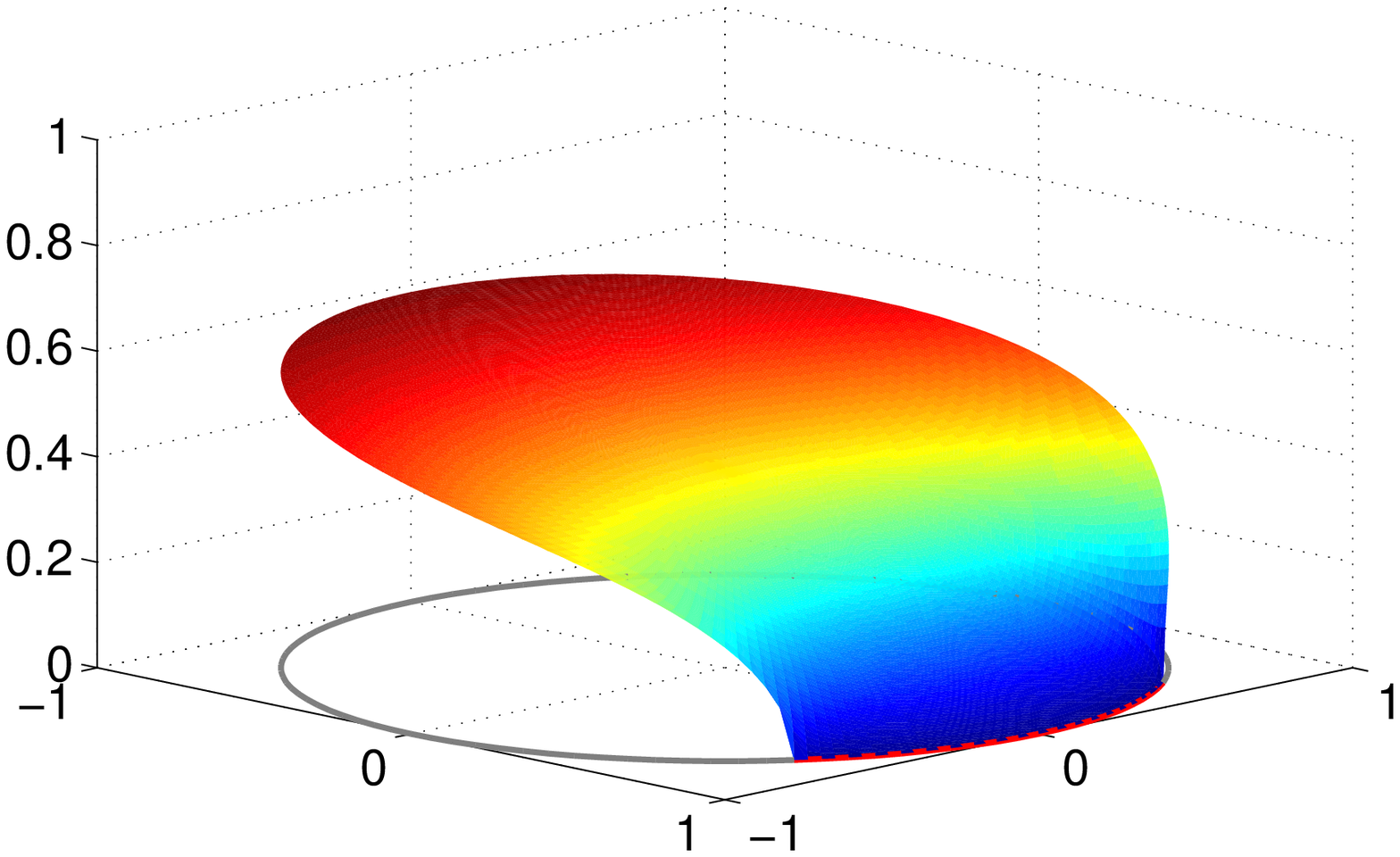} 
\includegraphics[width=85mm]{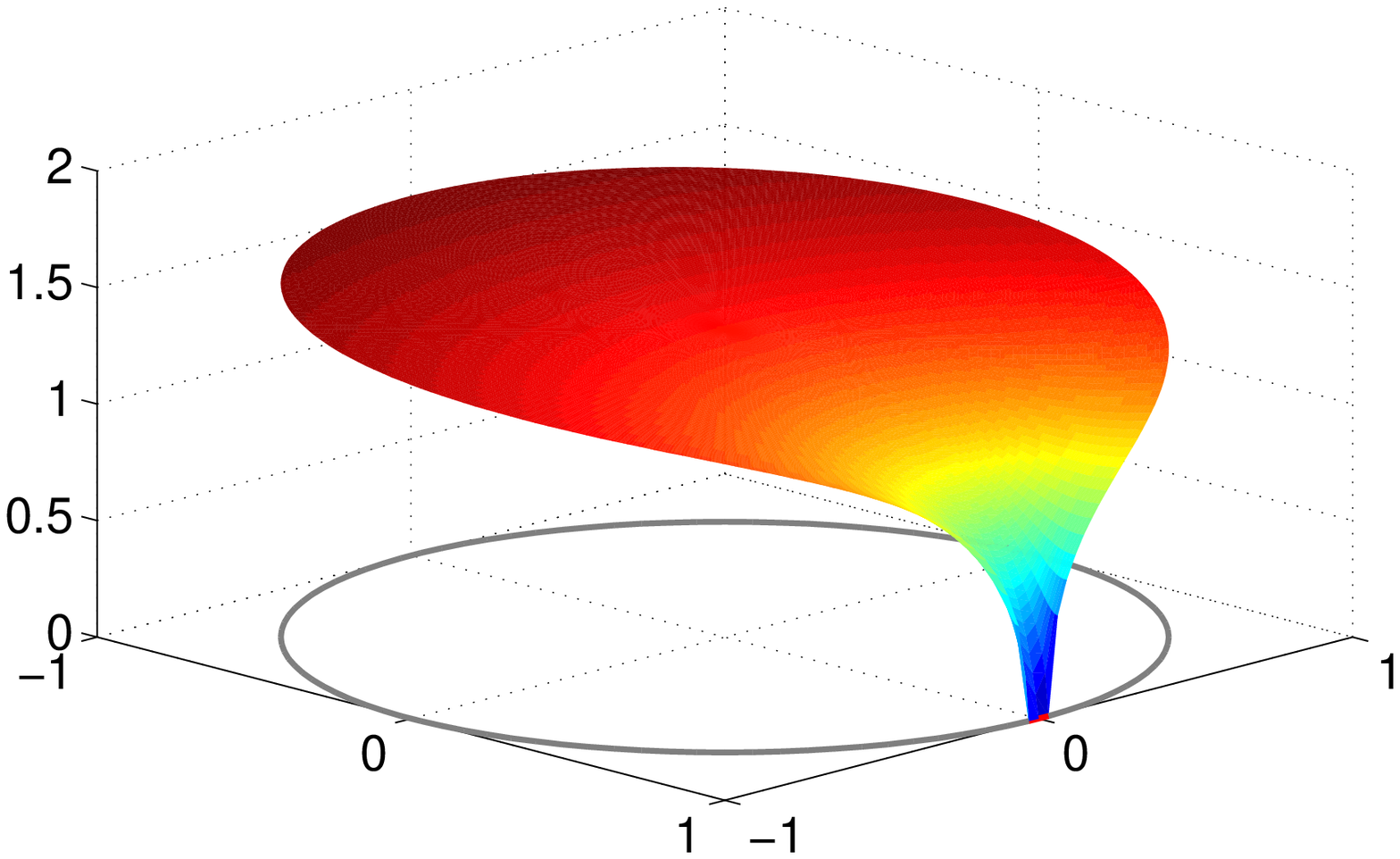} 
\end{center}
\caption{
The function $W_\omega(z)$ for $\omega = 0.2$ (top) and $\omega =
0.01$ (bottom).  It vanishes on the arc $(-\pi\omega,\pi\omega)$.}
\label{fig:Wtot}
\end{figure}

In order to get an explicit representation of $W_\omega(z_0)$, we
rewrite the Poisson kernel in Eq. (\ref{eq:F_Poisson}) as
\begin{equation}
\omega_{z_0}(\theta) = \frac{1}{2\pi} \biggl[1 + 2\sum\limits_{n=1}^\infty r_0^n \cos n(\theta-\theta_0)\biggr],
\end{equation}
so that
\begin{equation}
\frac12 \bigl[\omega_{z_0}(\theta) + \omega_{z_0}(-\theta)\bigr] = \frac{1}{2\pi} \biggl[1 + 2\sum\limits_{n=1}^\infty r_0^n \cos n\theta \cos n\theta_0\biggr].
\end{equation}
Integration of this function with respect to $\theta$ yields
\begin{equation}
\begin{split}
W_\omega(z_0) & = \frac{1}{\pi^2 \sqrt{2}} \int\limits_0^c d\theta \frac{\sin(\theta/2)}{\sqrt{\cos \theta - \cos c}}  \\
& \times \biggl[\theta + 2\sum\limits_{n=1}^\infty \frac{r_0^n}{n} \cos n \theta_0 \sin n\theta\biggr]. \\
\end{split}
\end{equation}
To proceed, we will use the identity (see \cite{Sneddon}, p. 59,
2.6.30)
\begin{equation}
\label{eq:kernel_Sneddon}
\frac{\sin (\theta/2)  \Theta(c-\theta)}{\sqrt{\cos \theta - \cos c}} = \frac{1}{\sqrt{2}}
\sum\limits_{k=1}^\infty \bigl(P_{k-1}(\cos c) - P_k(\cos c)\bigr) \sin k\theta ,
\end{equation}
where $\Theta(c-\theta)$ is the Heaviside step function.  Multiplying
this identity by $\theta$ and integrating over $\theta$ from $0$ to
$\pi$, we get
\begin{equation}
\begin{split}
& \int\limits_0^c d\theta \frac{\sin (\theta/2)}{\sqrt{\cos \theta - \cos c}} \, \theta \\
& = \frac{\pi}{\sqrt{2}} \sum\limits_{k=1}^\infty \bigl(P_{k-1}(\cos c) - P_k(\cos c)\bigr) \frac{(-1)^{k+1}}{k}  \\
& = - \pi \sqrt{2} \ln \cos(c/2). \\
\end{split}
\end{equation}
To prove the last relation, one can decompose $\ln(1+x)$ (with $x =
\cos c$) on Legendre polynomials as
\begin{equation*}
\begin{split}
\ln(1+x) & = \ln 2 - 1 + \sum\limits_{n=1}^{\infty} (-1)^{n+1} P_n(x) \frac{2n+1}{n(n+1)} \\
& =  \ln 2 + \sum\limits_{n=1}^{\infty} (-1)^{n+1} \frac{P_n(x) - P_{n-1}(x)}{n} . \\
\end{split}
\end{equation*}
Similarly, one multiplies Eq. (\ref{eq:kernel_Sneddon}) by $\sin
n\theta$ and integrates from $0$ to $\pi$ to get
\begin{equation}
\begin{split}
\int\limits_0^c d\theta \frac{\sin (\theta/2)}{\sqrt{\cos \theta - \cos c}}  \sin n \theta 
& = \frac{\pi \bigl(P_{n-1}(\cos c) - P_n(\cos c)\bigr)}{2\sqrt{2}}  . \\
\end{split}
\end{equation}
Combining these results, we find
\begin{equation}
\label{eq:Wc}
\begin{split}
W_\omega(z_0) &= - \frac{\ln \cos (c/2)}{\pi}  \\
& + \frac{1}{2\pi} \sum\limits_{n=1}^\infty \frac{P_{n-1}(\cos c) - P_n(\cos c)}{n} \, r_0^n \cos n\theta_0 .  \\
\end{split}
\end{equation}
Using Eq. (\ref{eq:c}) and $P_n(-x)=(-1)^n P_n(x)$, we obtain
\begin{equation}
\label{eq:Wc1}
\begin{split}
W_\omega(z_0) & = - \frac{\ln \sin (\pi\omega/2)}{\pi} \\
& - \sum\limits_{n=1}^\infty \frac{P_{n-1}(\cos \pi\omega) + P_n(\cos \pi\omega)}{2\pi n} \, r_0^n \cos n\theta_0 , \\
\end{split}
\end{equation}
where the arc was rotated back by replacing $\theta_0$ by
$\pi-\theta_0$.  Using the following identity (that we will prove
independently in \ref{sec:derivation})
\begin{equation}
\label{eq:Cagi_identity}
\begin{split}
& \sum\limits_{n=1}^\infty \frac{P_n(\cos \epsilon) + P_{n-1}(\cos \epsilon)}{2n}~  r^n \cos n\theta  \\
& = - \ln \left(\frac{|1 - re^{i\theta} + \sqrt{(1-re^{i(\theta-\epsilon)})(1-re^{i(\theta+\epsilon)})}|}{2}\right) , \\
\end{split}
\end{equation}
we find another representation
\begin{equation}
\label{eq:Womega}
W_\omega(z_0) = \frac{1}{\pi} \ln \left(\frac{|1-z_0 + \sqrt{(1-z_0e^{i\pi \omega})(1-z_0e^{-i\pi\omega})}|}{2\sin (\pi\omega/2)}\right).
\end{equation}
The behavior of this function is illustrated in Fig. \ref{fig:Wtot}.

Similarly, one can find the antisymmetric solution by using an
expansion over sine functions:
\begin{equation}
u_-(r,\theta) = \sum\limits_{n=1}^\infty a'_n r^n \sin n\theta ,
\end{equation}
with unknown coefficients $a'_n$ satisfying dual trigonometric
equations obtained from the boundary conditions.  However, since we
are only interested in $u(0,0)$, there is no need to solve this
equation.

Gathering the above results, we finally get
\begin{equation}
\label{eq:Tfinal0}
\begin{split}
\T(x_0) & = \tau(0) = v(0) + u_+(0,0) + u_-(0,0)   \\
&  = \int\limits_\D \frac{dz\, |\phi'_{x_0}(z)|^2}{D(\phi_{x_0}(z))}  \, \biggl(- \frac{\ln |z|}{2\pi} + W_\omega(z) \biggr)  , \\
\end{split}
\end{equation}
from which the change of the integration variable $z$ to $x =
\phi_{x_0}(z)$ yields
\begin{equation}
\label{eq:STfinal}
\T(x_0) = \int\limits_\Omega \frac{dx}{D(x)} \biggl(- \frac{\ln |\phi_{x_0}^{-1}(x)|}{2\pi} + W_\omega(\phi_{x_0}^{-1}(x)) \biggr) .
\end{equation}
Note that the factor $|\phi'_{x_0}(z)|^2$, which stands in
Eq. (\ref{eq:Tfinal0}) is the Jacobian of this change of variables:
$dx = dz\, |\phi'_{x_0}(z)|^2$.

In the formula (\ref{eq:STfinal}), the first integral that we denote as
\begin{equation}
\T_D(x_0) = - \int\limits_\Omega \frac{dx}{D(x)} \frac{\ln |\phi_{x_0}^{-1}(x)|}{2\pi} \, ,
\end{equation}
is the MFPT to the whole boundary, whereas the second term accounts
for eventual reflections on the remaining part of the boundary,
$\pa\backslash \Gamma$, until reaching the escape region $\Gamma$.

We recall that the harmonic measure $\omega = \omega_{x_0}(\Gamma)$
depends on the starting point $x_0$ and solves the following Dirichlet
boundary value problem in $\Omega$:
\begin{equation}
\label{eq:HM_def}
\left\{  \begin{array}{r l l}
\Delta \omega_{x_0}(\Gamma) & = 0  & x_0 \in \Omega, \\
 \omega_{x_0}(\Gamma) & = 1 & x_0 \in \Gamma,  \\
 \omega_{x_0}(\Gamma) & = 0 & x_0 \in \pa\backslash \Gamma  .\\  \end{array} \right.
\end{equation}
It can also be expressed in terms of the Green's function, or through
the conformal map by noting that the endpoints $x_1$ and $x_2$ of the
escape region $\Gamma$ (enumerated in the counterclockwise order) are
mapped back onto the points $e^{-i\pi\omega}$ and $e^{i\pi\omega}$ of
the circle of the unit disk.  In general, one has
\begin{equation}
x_1 = \phi_{x_0}(e^{i\alpha - i\pi\omega}),  \qquad
x_2 = \phi_{x_0}(e^{i\alpha + i\pi\omega}),
\end{equation}
where the factor $e^{i\alpha}$ aims at rotating the unit disk to put
the pre-image $\gamma$ onto the arc $(-\pi\omega,\pi\omega)$.  One
gets therefore
\begin{equation}
\label{eq:alpha}
\begin{split}
e^{2i\pi\omega} & = \phi_{x_0}^{-1}(x_1)/\phi_{x_0}^{-1}(x_2) , \\
e^{2i\alpha} & = \phi_{x_0}^{-1}(x_1) \, \phi_{x_0}^{-1}(x_2) , \\
\end{split}
\end{equation}
from which both $\omega$ and $\alpha$ can be determined.

Although the MFPT and the harmonic measure are both formulated as
boundary value problems for the Laplace operator, their relation is
not so intuitive.  In fact, the harmonic measure $\omega =
\omega_{x_0}(\Gamma)$ is the probability that the first arrival of
Brownian motion on the boundary occurs at the escape region $\Gamma$.
In other words, the harmonic measure characterizes the
``accessibility'' of the escape region $\Gamma$ in ``competition''
with the remaining part of the boundary $\pa\backslash \Gamma$
\cite{Grebenkov05}.  This competition is known as diffusion screening
\cite{Meakin85,Sapoval94}, the concept that has found numerous
implications for heterogeneous catalysis \cite{Filoche08}, fluid flow
in rough channels \cite{Andrade07,Bazant16} and transport phenomena in
biological systems \cite{Felici03,Grebenkov05b,Gill11,Maoy12}.  In
turn, only the escape region is absorbing in the MFPT problem while
the remaining boundary is reflecting.  In particular, after hitting
the reflecting boundary, the particle bounces back into the domain and
continues to diffuse.  The statistics of Brownian trajectories in
these two settings are thus very different, that makes the relation
(\ref{eq:STfinal}) particularly intricate.

At the end of this section, we outline the main mathematical facts
that we used for derivation: (i) the infinitesimal generator of the
Brownian motion is the Laplace operator; (ii) there exists a conformal
mapping from the unit disk to any simply connected planar domain
(Riemann's theorem); (iii) the conformal map preserves angles; and
(iv) the Poisson equation with mixed Dirichlet-Neumann boundary
condition can be analytically solved in the disk.  These facts
highlight the intrinsic orientation of the proposed approach to planar
Brownian motion.  In particular, extensions to other stochastic
processes (governed by a general elliptic or Fokker-Planck operator)
or to higher dimensions remain unknown.

\section{Numerical implementation and validation}
\label{sec:validation}

Practical implementation of the exact formula (\ref{eq:STfinal})
involves two numerical steps: computation of the conformal map and
integration.

(i) For a given polygonal domain, the first step can be realized
either by a Schwarz-Christoffel formula \cite{Driscoll}, or by a
``zipper'' algorithm \cite{Marshall07}.  Once a conformal map
$\phi_{x_0}(z)$ is constructed for a starting point $x_0$, the
M\"obius transform,
\begin{equation}
\label{eq:Mobius}
M_{z_0}(z) = \frac{z_0 - z}{1 - \bar{z}_0 z} \,,
\end{equation}
yields the conformal map $\phi_{x'_0}(z)$ for another starting point
$x'_0$:
\begin{equation}
\label{eq:Mtransform}
\phi_{x'_0}(z) = \phi_{x_0}\bigl(M_{\phi^{-1}_{x_0}(x'_0)}(z) \bigr) .
\end{equation}
In other words, the numerical construction of the conformal map is
needed only for one starting point.  Note that the transformation
(\ref{eq:Mtransform}) can also be helpful to investigate the
dependence of the MFPT on the starting point.

(ii) The numerical integration involves meshing of the domain and
quadratures.  The MFPT can be computed through either of two
equivalent representations (\ref{eq:Tfinal0}, \ref{eq:STfinal}).  While
the integration over the unit disk in Eq. (\ref{eq:Tfinal0})
facilitates meshing, it requires an accurate numerical treatment of
integrable singularities of the derivative of the conformal map near
pre-vertices (the factor $|\phi'_{x_0}(z)|^2$).  For this reason, we
use the other option, in which the domain $\Omega$ is meshed by
triangles, the integrand function in Eq. (\ref{eq:STfinal}) is
evaluated at the vertices of these triangles and then summed to
approximate the integral.  We checked the accuracy of computation by
doubling the number of triangles.

Although both numerical steps are well documented and controlled, it
is instructive to illustrate their accuracy on two simple domains:
disk and rectangle.

\subsection{Disk}

\begin{figure}
\begin{center}
\includegraphics[width=85mm]{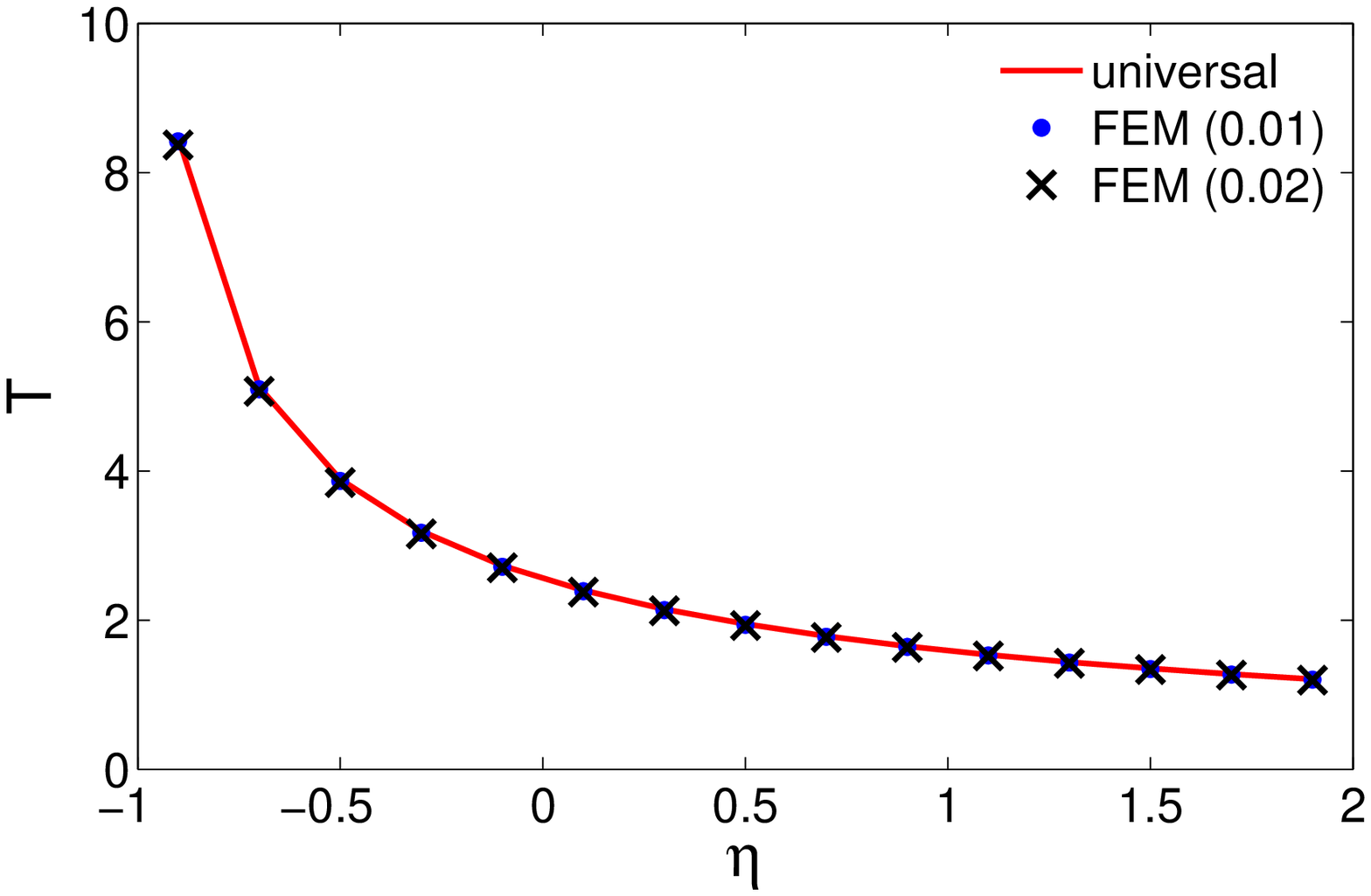} 
\end{center}
\caption{
The MFPT $\T(0)$ through the escape arc $(\pi-\epsilon,\pi+\epsilon)$
(with $\epsilon = 0.2$) on the boundary of the unit disk as a function
of the gradient $\eta$ of the linearly varying diffusion coefficient
along the radial coordinate: $D(x_0) = 1 + \eta |x_0|$.  The universal
formula (\ref{eq:STfinal}, solid line) is compared to FEM solutions of
Eqs. (\ref{eq:SFP_Omega}) with two mesh sizes: 0.01 (circles) and 0.02
(crosses).}
\label{fig:disk_Dr}
\end{figure}

We first consider the MFPT through the escape arc
$(\pi-\epsilon,\pi+\epsilon)$ on the boundary of the unit disk with a
linearly varying diffusion coefficient along the radial coordinate:
$D(x_0) = D(1 + \eta |x_0|)$, with a gradient $\eta > -1$.  In the
special case $\eta = 0$, one recovers the uniform diffusion
coefficient, for which the explicit formula (\ref{eq:Tdisk}) can be
used (see \ref{sec:disk} for details).  For other cases ($\eta \ne
0$), we resort to a numerical solution of the original mixed boundary
value problem (\ref{eq:SFP_Omega}) by a finite element method (FEM)
implemented in the Matlab PDE toolbox.  Our universal formula
(\ref{eq:STfinal}) is compared to a FEM numerical solution in
Fig. \ref{fig:disk_Dr}.  To control the quality of the FEM solution,
we provide the numerical results for two mesh sizes.  One observes an
excellent agreement between the universal formula and both numerical
solutions.  Since the conformal map is trivial for this domain, the
numerical implementation of the universal formula is much faster and
much more accurate than that of a FEM.

\subsection{Rectangle}

\begin{figure}
\begin{center}
\includegraphics[width=85mm]{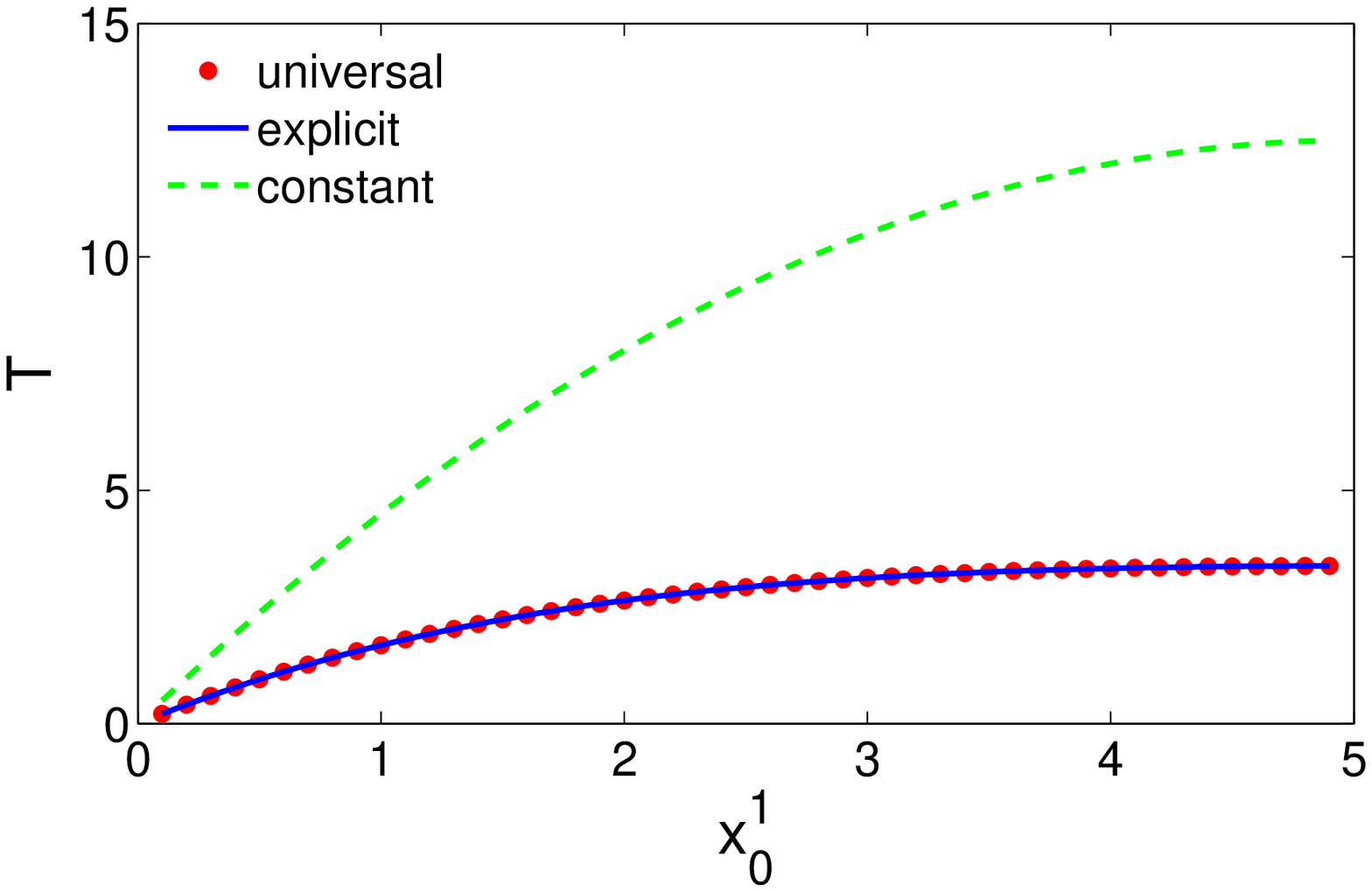} 
\end{center}
\caption{
The MFPT $\T(x_0)$ through the left edge of the rectangle $[0,L]\times
[0,h]$ for space-dependent diffusion coefficient $D(x_0)$ in
Eq. (\ref{eq:Dx_sin}), with $L = 5$, $h = 1$, $x_0^2 = 0.5$, $\beta =
-0.9$, $m = 0$, and $D = 1$.  The universal formula (\ref{eq:STfinal},
circles) is compared to the explicit solution (\ref{eq:T_Dx_sin},
solid line) available exclusively for this setting.  Dashed line shows
the first term, $(Lx_0^1 - \frac12 (x_0^1)^2)/D$, corresponding to the
constant diffusion coefficient.}
\label{fig:rectangle_Dx}
\end{figure}

We next consider the MFPT to the left edge of the rectangle
$[0,L]\times [0,h]$ with the diffusion coefficient $D(x_0)$ which
depends only on the horizontal coordinate $x_0^1$.  In this particular
setting, the MFPT does not depend on the vertical coordinate $x_0^2$,
and the remaining one-dimensional problem can be solved exactly:
\begin{equation}
\T(x_0) = \int\limits_0^L dx^1 \frac{G_1(x_0^1,x^1)}{D(x^1)} \,,
\end{equation}
where $G_1(x_0^1,x^1)$ is the Green function for the interval $[0,L]$
with Dirichlet and Neumann boundary conditions at endpoints $0$ and
$L$, respectively:
\begin{equation}
G_1(x_0^1,x^1) = 2L \sum\limits_{n=0}^\infty \frac{\sin\bigl(\pi(n+\frac12)\frac{x_0^1}{L}\bigr) \, 
\sin\bigl(\pi(n+\frac 12)\frac{x^1}{L}\bigr)}{\pi^2 (n+\frac 12)^2} .
\end{equation}
For illustrative purposes, we choose a particular spatial dependence
\begin{equation}
\label{eq:Dx_sin}
D(x^1) = \frac{D}{1 + \beta \sin\bigl(\pi(m+\frac12) \frac{x^1}{L}\bigr)}
\end{equation} 
in order to get a simple explicit solution:
\begin{equation}
\label{eq:T_Dx_sin}
\T(x_0) = \frac{L x_0^1 - \frac12 (x_0^1)^2}{D} + \frac{\beta L^2}{D} \, \frac{\sin\bigl(\pi(m+\frac12) \frac{x_0^1}{L}\bigr)}{\pi^2 (m+\frac12)^2} \,,
\end{equation}
where $m$ is an integer and $|\beta| < 1$.  Here the first term is the
MFPT that would be obtained for a constant diffusivity, while the
second term results from periodic fluctuations of the chosen spatial
dependence in $D(x^1)$.  Figure \ref{fig:rectangle_Dx} illustrates the
high accuracy of the numerical computation by the universal formula
(\ref{eq:STfinal}).  Although this example may look too simplistic,
numerical computation of conformal maps is known to be challenging for
elongated shapes because of the crowding phenomenon \cite{Driscoll}.
In spite of this potential difficulty, the maximal relative error of
our numerical implementation of the universal formula in this example
is below $0.003\%$.

\subsection{Comparison to other numerical techniques}

Since practical implementations of the exact solution
(\ref{eq:STfinal}) involve numerical steps, one may wonder how
efficient this approach is in comparison to conventional numerical
techniques for solving the boundary value problem (\ref{eq:SFP_Omega})
such as finite element or finite difference methods or Monte Carlo
simulations \cite{Cheviakov12}.  Although a systematic comparison
between different techniques is beyond the scope of this letter, we
outline one of the major advantages of the present approach from the
numerical point of view.  In general, conventional numerical
techniques suffer in the narrow escape limit.  Indeed, both finite
element and finite difference methods would require very fine meshes
to accurately treat the mixed boundary condition near a small escape
region.  Similarly, Monte Carlo simulations would be slowed down as
longer trajectories need to be generated to access larger FPTs, while
the number of these trajectories has be increased to compensate for
higher dispersion of FPTs in the narrow escape limit.  In contrast,
the numerical implementation of the exact solution (\ref{eq:STfinal})
is expected to be less sensitive to the escape region size: (i) the
computation of the conformal map is independent of boundary conditions
and of the escape region size, and (ii) numerical integration of a
smooth integrand function in Eq. (\ref{eq:STfinal}) does not require
very fine meshes and can be further improved by using high order
quadratures.  In other words, the numerical advantage of the proposed
approach results from the natural ``representation'' of a confining
domain by the conformal map and from the incorporation of the mixed
boundary condition through the explicit function $W_\omega(z)$.
Moreover, the narrow escape limit is particularly favorable for the
present approach since the asymptotic formula (\ref{eq:STfinalA}) from
\ref{sec:asymptotic} becomes very accurate so that the leading
logarithmic and constant terms can be enough for an accurate
evaluation of the MFPT.

\section{Analytical results for disk and rectangle}
\label{sec:analytical}

To illustrate the use of the main formula (\ref{eq:STfinal}), we
consider the MFPT in two basic domains, disk and rectangle, with a
constant diffusion coefficient.

\subsection{Disk}
\label{sec:disk}

For the unit disk with an escape arc $(-\epsilon,\epsilon)$, Singer
{\it et al.}  provided the exact solution for arbitrary $\epsilon$ in
terms of an infinite series with coefficients in the form of integrals
\cite{Singer06b} that were later reduced to Legendre polynomials
\cite{Rupprecht15}:
\begin{equation}
\label{eq:Tx0_disk}
\begin{split}
\T(x_0) & = \frac{1}{D} \biggl( \frac{1-r_0^2}{4} - \ln \sin(\epsilon/2) \\
& - \sum\limits_{n=1}^\infty \frac{P_n(\cos \epsilon) + P_{n-1}(\cos \epsilon)}{2n}  r_0^n \cos n\theta_0 \biggr), \\
\end{split}
\end{equation}
with $x_0 = r_0 e^{i\theta_0}$.  A simpler explicit formula was
obtained by Caginalp and Chen \cite{Caginalp12}:
\begin{equation}
\label{eq:Tdisk}
\begin{split}
\T(x_0) & = \frac{1}{D} \biggl[ \frac{1-r_0^2}{4}  \\
& + \ln \biggl(\frac{|1 - x_0 + \sqrt{(1-x_0e^{-i\epsilon})(1-x_0e^{i\epsilon})}|}{2\sin(\epsilon/2)}\biggr) \biggr]. \\
\end{split}
\end{equation}
The comparison of these two relations implies the identity
(\ref{eq:Cagi_identity}) that we derive independently in
\ref{sec:derivation}.

Although the solution for the disk is known, it is instructive to
recover it from the general formula (\ref{eq:Tfinal0}).  Conformal
mapping of the unit disk $\D$ onto itself is realized by a family of
linear fractional transformations
\begin{equation}
\phi(z) = e^{i\alpha} \frac{z - w}{1 - \bar{w} z} \,, 
\end{equation}
where $\alpha$ is a real number and $w$ is a point in $\D$.  Imposing
the mapping of the origin of the disk onto $x_0 \in \Omega$ ($=\D$),
we set
\begin{equation}
\label{eq:phiz_disk}
\phi_{x_0}(z) = \frac{x_0 - z \, e^{i\alpha}}{1 - z \, \bar{x}_0 \, e^{i\alpha}} \,,  
\end{equation}
where $\alpha$ is a real parameter which determines an appropriate
rotation of the disk (in $z$ coordinates) to ensure the symmetry of
the escape arc, $(-\pi \omega,\pi \omega)$.  One gets thus
\begin{equation}
\label{eq:phip}
|\phi'_{x_0}(z)| = \frac{1-|x_0|^2}{1 - z \bar{x}_0 e^{i\alpha} - \bar{z} x_0 e^{-i\alpha} + |z|^2 |x_0|^2} \,,
\end{equation}
while the inverse mapping is
\begin{equation}
\label{eq:phiinv}
\phi_{x_0}^{-1}(x) = \frac{x_0 - x}{1 - x \bar{x}_0} e^{-i\alpha} .
\end{equation}

Relating the escape region $\Gamma = (-\epsilon,\epsilon)$ and its
pre-image $\gamma = (-\pi \omega,\pi\omega)$ by the conformal map
(\ref{eq:phiz_disk}),
\begin{equation}
e^{i\epsilon} = \phi_{x_0}(e^{i\pi\omega}) , \qquad  e^{-i\epsilon} = \phi_{x_0}(e^{-i\pi\omega}) ,
\end{equation}
one finds $\omega$ and $\alpha$
\begin{equation}
\begin{split}
\omega & = - \frac{i}{2\pi} \ln \left(\frac{e^{i\epsilon} - x_0}{\bar{x}_0 e^{i\epsilon} - 1} \frac{\bar{x}_0 e^{-i\epsilon} - 1}{e^{-i\epsilon} - x_0}\right), \\
\alpha & = -i \ln \left(- \sqrt{\frac{e^{i\epsilon} - x_0}{\bar{x}_0 e^{i\epsilon} - 1} \frac{e^{-i\epsilon} - x_0}{\bar{x}_0 e^{-i\epsilon} - 1}}\right). \\
\end{split}
\end{equation}
These relations can also be re-written as
\begin{equation}
\label{eq:omega_disk}
\omega = \omega_{x_0}(\Gamma) = \begin{cases} \frac{1}{\pi} \arctan(\eta)  \hskip 13mm (\eta > 0), \cr
1 + \frac{1}{\pi} \arctan(\eta)  \qquad (\eta < 0) ,  \end{cases}
\end{equation}
with
\begin{equation}
\eta = \frac{(1-r_0^2)\sin \epsilon}{(1+r_0^2) \cos \epsilon - 2r_0 \cos\theta_0} \,,
\end{equation}
and
\begin{equation}
\label{eq:alpha_disk}
\alpha = \arctan \left(\frac{2r_0 \sin \theta_0 (r_0 \cos\theta_0 - \cos \epsilon)}{1-r_0^2 + 2r_0 \cos\theta_0 (r_0 \cos\theta_0 - \cos \epsilon)}\right).
\end{equation}
Note that the harmonic measure $\omega$ could alternatively be
determined by integrating the Poisson kernel (\ref{eq:Poisson}).

Substituting (\ref{eq:phiinv}) into the first term of
Eq. (\ref{eq:Tfinal0}) and computing the integral in polar
coordinates, one retrieves the classical MFPT to the unit circle:
\begin{equation}
\T_D(x_0) = \frac{1-r_0^2}{4D} ,
\end{equation}
where we used the identity (see \cite{Gradshteyn}, p. 541)
\begin{equation}
\int\limits_0^{2\pi} d\theta ~ \ln(1 + 2r \cos \theta + r^2) = 0  \qquad \mbox{for}~ |r| \leq 1 .
\end{equation}
Using the series representation (\ref{eq:Wc1}), one computes the
second contribution in Eq. (\ref{eq:Tfinal0}) by integrating term by
term
\begin{equation*}
\begin{split}
& \int\limits_\D \frac{dz}{D} \, W_\omega(z) \, |\phi'_{x_0}(z)|^2 = \frac{-1}{2\pi D} \sum\limits_{n=1}^\infty c_n \\
& \times  \int\limits_0^1 dr \, r \int\limits_0^{2\pi} d\theta \frac{(1-r_0^2)^2 \, r^n \cos n\theta}{(1 - 2r r_0 \cos(\theta-\theta_0+\alpha) + r_0^2 r^2)^2} \\
& = \frac{-1}{D} \sum\limits_{n=1}^\infty c_n r_0^n \cos n(\theta_0-\alpha) ,  \\
\end{split}
\end{equation*}
with
\begin{equation}
c_n = \frac{P_{n-1}(\cos \pi\omega) + P_n(\cos \pi\omega)}{2n} ,
\end{equation}
where we used the identity
\begin{eqnarray*}
\int\limits_0^1 dr  \int\limits_0^{2\pi} \frac{r^{n+1} \cos n\theta \, d\theta}{(1-2ar\cos(\theta-\theta_0) + a^2r^2)^2} 
&=& \frac{\pi a^n \cos n\theta_0}{(1-a^2)^2} \,.
\end{eqnarray*}
Combining these results, one gets
\begin{widetext}
\begin{equation}
\label{eq:T_disk_new}
\begin{split}
\T(x_0) & = \frac{1}{D} \biggl[\frac{1-r_0^2}{4} - \frac{1}{\pi} \ln \sin(\pi\omega/2) -
\sum\limits_{n=1}^\infty \frac{P_{n-1}(\cos \pi\omega) + P_n(\cos \pi\omega)}{2n}  r_0^n \cos n(\theta_0-\alpha) \biggr] \\
& = \frac{1}{D} \biggl[\frac{1-r_0^2}{4} + \ln \biggl(\frac{|1-r_0 e^{i(\theta_0-\alpha)} + 
\sqrt{(1-r_0 e^{i(\theta_0-\alpha+\pi\omega)})(1-r_0 e^{i(\theta_0-\alpha-\pi\omega)})}|}{2 \sin(\pi\omega/2)} \biggr) \biggr], \\
\end{split}
\end{equation}
\end{widetext}
where we used the identity (\ref{eq:Cagi_identity}).  Comparing
Eqs. (\ref{eq:Tdisk}, \ref{eq:T_disk_new}), one gets the following
relation for the unit disk:
\begin{equation}
W_\ve(x_0) = W_{\omega}(x_0 e^{-i\alpha}),
\end{equation}
where $\omega$ and $\alpha$ are related to $x_0$ according to
Eqs. (\ref{eq:omega_disk}, \ref{eq:alpha_disk}), and $\ve =
|\Gamma|/|\pa| = \epsilon/\pi$.

\subsection{Thin rectangle}

We consider now the MFPT from a thin rectangle $\Omega = [0,L] \times
[0,h]$ through its left edge: $\Gamma = \{0\} \times [0,h]$.  The
exact solution of this problem is simply
\begin{equation}
\label{eq:Trectangle}
\T(x_0) = \frac{2Lx_0^1 - [x_0^1]^2}{2D}  ,
\end{equation}
which does not depend on $x_0^2$ and $h$ (in this subsection, we use
the Cartesian coordinates, $x_0 = (x_0^1,x_0^2)$, instead of polar
coordinates or complex numbers).  One can see that the MFPT is not
determined by the normalized perimeter $\ve = \frac{h}{2(L+h)}$, even
if the latter is very small.

The harmonic measure $\omega_{x_0}(\Gamma)$ is obtained by solving
Eq. (\ref{eq:HM_def}):
\begin{equation}
\omega_{x_0} = \sum\limits_{n=1}^\infty \frac{2(1-(-1)^n)}{\pi n} \sin (\pi n x_0^2/h) \frac{\sinh(\pi n(L-x_0^1)/h)}{\sinh(\pi nL/h)} .
\end{equation}
Setting the starting point on the horizontal line at the middle,
$x_0^2 = h/2$, and omitting exponentially small terms with
$n=2,3,\ldots$ (for $h\ll L$), one gets $\omega_{(x_0^1,h/2)} \simeq
\frac{4}{\pi} e^{-\pi x_0^1/h}$, from which Eq. (\ref{eq:STfinalA})
yields the leading term
\begin{equation}
\label{eq:T_rect}
\T(x_0^1,h/2) \simeq - \frac{Lh}{\pi D} \ln \left(\omega_{(x_0^1,h/2)}\right) \simeq \frac{Lx_0^1}{D}  , 
\end{equation}
in agreement with the exact solution (\ref{eq:Trectangle}).  The
missing term $-\frac12 [x_0^1]^2/D$ from Eq. (\ref{eq:Trectangle})
is related to the presence of the reflecting edge at $x^1 = L$ which
is accounted for by the remaining terms in Eq. (\ref{eq:STfinalA}).  If
the starting point $x_0$ was located near a long edge (i.e., if
$x_0^2$ was close to $0$ or $h$), one would get an extra term, $-
\frac{Lh}{\pi D} \ln \sin(\pi x_0^2/h)$ but its contribution would
be compensated by the remaining terms in Eq. (\ref{eq:STfinalA}) that
we ignored here.  Note that the MFPT to the whole boundary,
$\T_D(x_0)$, can be computed exactly but it vanishes as $h\to 0$.

Now we consider the MFPT from the same rectangle but through any edge
except the right one.  Here the Poisson equation, $\Delta \T(x_0) =
-1/D$, is completed by boundary conditions: $\T(x_0^1,0) =
\T(x_0^1,h) = \T(0,x_0^2) = 0$ (escape region) and $\partial
\T/\partial x_0^1 = 0$ at $x_0^1 = L$ (reflecting region).  One gets an
explicit solution
\begin{equation}
\begin{split}
\T(x_0) & = \frac{x_0^2(h-x_0^2)}{2D} - \frac{2h^2}{D} \sum\limits_{n=1}^\infty \frac{1-(-1)^n}{\pi^3 n^3} \\
& \times  \sin(\pi nx_0^2/h) \frac{\cosh(\pi n(L-x_0^1)/h)}{\cosh(\pi nL/h)} . \\
\end{split}
\end{equation}
When the starting point $x_0$ is not close to the left or right edges
(i.e., $h \ll x_0^1$ and $h \ll L-x_0^1$), the contribution from the
second term is exponentially small (of the order of $e^{-\pi
x_0^1/h}$), and the MFPT is determined by the first term (describing
the one-dimensional problem along the vertical coordinate).  In the
limit $x_0^1\to 0$, one gets (for $L \gg h$)
\begin{equation}
\label{eq:T_rect2}
\T(x_0) \simeq  x_0^1 \, \frac{2h}{D} \sum\limits_{n=1}^\infty \frac{1-(-1)^n}{\pi^2 n^2} \sin(\pi nx_0^2/h) .  
\end{equation}
For instance, setting $x_0^2 = h/2$, the sum is evaluated numerically,
yielding $\T(x_0) \simeq 0.37 \, x_0^1 h/D$.  This relation is
similar to Eq. (\ref{eq:T_rect}), with the length $L$ of the rectangle
being replaced by its width $h$.
%
%
In both relations (\ref{eq:T_rect}, \ref{eq:T_rect2}), the MFPT does
not scale with the area of the domain.  Moreover, in the latter case,
one can take the limit $L\to \infty$ and consider an infinite
half-stripe (of infinite area) for which the MFPT remains finite.

\section{Asymptotic behavior}
\label{sec:asymptotic}

When the escape region $\Gamma$ is the whole boundary (no reflecting
part), the harmonic measure $\omega$ is equal $1$, the function
$W_\omega(z)$ vanishes, and one recovers the conventional solution
$\T_D(x_0)$ for the Dirichlet problem, as expected.  

In this Section, we focus on the more interesting limit $\omega\to 0$.
Using
\begin{equation}
P_n(\cos(\pi \omega)) = 1 - \frac{n(n+1)}{4} \pi^2 \omega^2 + O(\omega^4), 
\end{equation}
we get
\begin{equation}
\label{eq:Womega_A}
\begin{split}
W_\omega(z) & = - \frac{\ln \omega}{\pi} + \biggl(\frac{\ln (2/\pi)}{\pi} + V_0(z)\biggr) \\
& + \omega^2 \biggl(\frac{\pi}{24} - V_2(z)\biggr) + O(\omega^4), \\
\end{split}
\end{equation}
where
%
%
\begin{widetext}
\begin{equation}
\begin{split}
V_0(z) & = - \frac{1}{2\pi} \sum\limits_{n=1}^\infty \frac{2}{n} \, r^n \cos n\theta 
 = \frac{1}{2\pi} \ln \bigl(1 - 2r \cos \theta + r^2\bigr) = \frac{1}{2\pi} \ln \bigl((1-z)(1-\bar z)\bigr) = \frac{1}{\pi} \ln |1-z| ,\\
V_2(z) & = - \frac{\pi}{4} \sum\limits_{n=1}^\infty  n \, r^n \cos n\theta 
 = \frac{\pi r}{4} \frac{2r - (1+r^2)\cos \theta}{(1-2r\cos\theta + r^2)^2}  
 = - \frac{\pi}{8} \biggl[\frac{z}{(1-z)^2} + \frac{\bar{z}}{(1-\bar{z})^2}\biggr] 
= \frac{\pi}{8} \biggl(\frac{(1+|z|^2)}{|1-z|^2} - \frac{(1-|z|^2)^2}{|1-z|^4}\biggr). \\
\end{split}
\end{equation}
\end{widetext}
Similarly, one can evaluate higher-order terms in
Eq. (\ref{eq:Womega_A}).  Substituting the expansion
(\ref{eq:Womega_A}) into Eq. (\ref{eq:STfinal}), one deduces the
asymptotic relation
\begin{widetext}
\begin{equation}
\label{eq:STfinalA}
\begin{split}
\T(x_0) & = -\frac{|\Omega|}{\pi D_h} \ln \omega + \biggl(\frac{|\Omega| \ln(2/\pi)}{\pi D_h}  + \frac{1}{2\pi} 
\int\limits_\Omega \frac{dx}{D(x)} \ln \biggl(\frac{|1-\phi^{-1}_{x_0}(x)|^2}{|\phi^{-1}_{x_0}(x)|} \biggr)\biggr)
+ \omega^2 \biggl(\frac{\pi |\Omega|}{24 D_h} - \int\limits_\Omega dx \frac{V_2(\phi^{-1}_{x_0}(x))}{D(x)} \biggr) + O(\omega^4) , \\
\end{split}
\end{equation}
\end{widetext}
where
\begin{equation}
\frac{1}{D_h} = \frac{1}{|\Omega|} \int\limits_\Omega \frac{dx}{D(x)} .
\end{equation}

We emphasize that the smallness of the harmonic measure $\omega =
\omega_{x_0}(\Gamma)$ is not related to the smallness of the escape
region $\Gamma$.  In fact, the harmonic measure characterizes the
``accessibility'' of $\Gamma$ by Brownian motion starting from $x_0$.
The escape region can be very small but if $x_0$ lies close to
$\Gamma$, the harmonic measure is close to $1$ (e.g., $\omega = 1$
when $x_0\in \Gamma$).  On the opposite, the escape region can be
large but almost ``inaccessible'' from $x_0$ due to diffusion
screening, in which case the harmonic measure is small (e.g., $\omega
= 0$ when $x_0 \in \pa\backslash \Gamma$).

\section{Generation of irregular domains}

\begin{figure}
\begin{center}
\includegraphics[width=55mm]{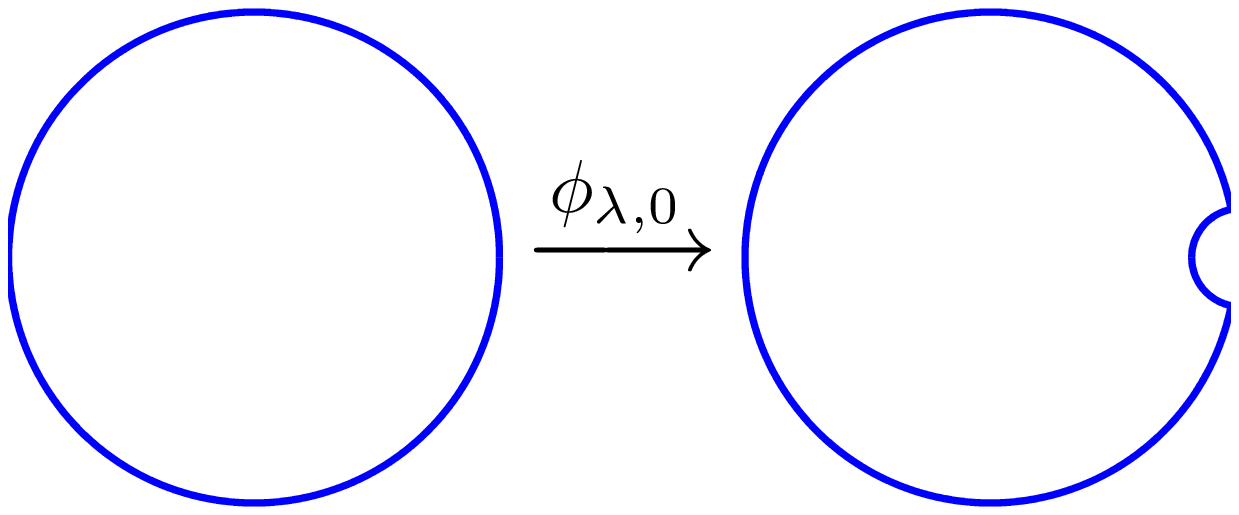} 
\end{center}
\caption{
The function $\varphi_{\lambda,0}(z)$ maps the unit disk onto the unit
disk without a semi-circular region of radius $\lambda = 0.2$. }
\label{fig:conformal_phi}
\end{figure}

As discussed in \ref{sec:validation}, the unit disk can be mapped onto
a given polygonal domain by various numerical tools such as a
Schwarz-Christoffel transformation \cite{Driscoll} or ``zipper''
algorithm \cite{Marshall07}.  It is also possible to create irregular
domains by means of iterative conformal maps.  For an illustrative
purpose, we choose this last option and adopt the Hastings-Levitov
algorithm for Laplacian growth \cite{Hastings98,Davidovitch99}.  This
algorithm is based on a conformal map that creates a circular ``bump''
on the unit disk.  Repeating this map iteratively with random bump
locations and appropriate rescaling of bump sizes, one can grow
DLA-like clusters and study the harmonic measure on its surface.
Since this algorithm maps the exterior of the unit disk onto the
exterior of the cluster, it is not directly applicable for our
purposes as we need a map from the unit disk onto the interior of a
bounded domain.  Inspired by this algorithm, we consider another basic
mapping, from the unit disk onto the unit disk without a nearly
semi-circular region:
\begin{equation}
\varphi_{\lambda,0}(z) = \frac{\sqrt{1+\lambda^2}}{2} \left(1 + z - \sqrt{1-z} \sqrt{\frac{1+z^2 - 2z \frac{1-\lambda^2}{1+\lambda^2}}{1-z}}\right),
\end{equation}
where $\lambda$ is close to the radius of the removed region (see
Fig. \ref{fig:conformal_phi}).  After $n$ iterations, the conformal
map reads
\begin{equation}
\label{eq:Phin}
\Phi^{(n)}(z) = \varphi_{\lambda_1,\theta_1}(\varphi_{\lambda_2,\theta_2}(\cdots \varphi_{\lambda_n,\theta_n}(z)\cdots)) ,
\end{equation}
where the map $\varphi_{\lambda,\theta}(z)$ specifies the location
$\theta$ of the removed region:
\begin{equation}
\varphi_{\lambda,\theta}(z) = e^{i\theta} \varphi_{\lambda,0}(e^{-i\theta}z) .
\end{equation}
Both locations $\theta_k$ and sizes $\lambda_k$ can in general be
chosen arbitrarily.  To reduce size distortions due to conformal maps
and keep physical sizes of all removed regions of the same order
$\lambda_0$, we set $\lambda_n =
\lambda_0/|\Phi^{(n-1)'}(e^{i\theta_n})|$, where $\Phi^{(n-1)'}$ is
the derivative of the map at the previous step $n-1$ which is
expressed through the explicitly computable derivatives of
$\varphi_{\lambda_k,\theta_k}$.  Note that $\Phi^{(n)}(z)$ maps $0$ to
$0$.  Once the conformal map is constructed (i.e., the sets
$\lambda_1,\ldots,\lambda_n$ and $\theta_1,\ldots,\theta_n$ are chosen
or determined), one can apply the M\"obius transformation to ensure
the mapping from $0$ to a given point $x_0$.  In fact, it is enough to
replace the argument $z$ in Eq. (\ref{eq:Phin}) by its M\"obius
transform $M_{z_0}(z)$ given in Eq. (\ref{eq:Mobius}), where $z_0$ is
the pre-image of the point $x_0$,
\begin{equation}
z_0 = [\Phi^{(n)}]^{-1}(x_0), 
\end{equation}
and this inverse conformal map is obtained as
\begin{equation}
[\Phi^{(n)}]^{-1}(x) = \varphi^{-1}_{\lambda_n,\theta_n}\bigl(\varphi^{-1}_{\lambda_{n-1},\theta_{n-1}}\bigl(\cdots 
\varphi^{-1}_{\lambda_1,\theta_1}(x)\cdots\bigr)\bigr) ,
\end{equation}
where
\begin{equation}
\varphi^{-1}_{\lambda,\theta}(x) = e^{i\theta} \varphi^{-1}_{\lambda,0}(e^{-i\theta}x) ,
\end{equation}
and
\begin{equation}
\varphi^{-1}_{\lambda,0}(x) = x \frac{x - \sqrt{1+\lambda^2}}{x \sqrt{1+\lambda^2} - 1} \,.
\end{equation}
Combining these steps, one gets
\begin{equation}
\begin{split}
\phi_{x_0}(z) & = \Phi^{(n)}(M_{z_0}(e^{i\alpha} z))  , \\
\phi^{-1}_{x_0}(x) & = e^{-i\alpha} M^{-1}_{z_0}([\Phi^{(n)}]^{-1}(x)), \\
\end{split}
\end{equation}
where the factor $e^{i\alpha}$ depends on the escape region $\Gamma$
and rotates the disk to ensure that the pre-image of $\Gamma$ is
$(-\pi\omega,\pi\omega)$, see Eq. (\ref{eq:alpha}).  The great
advantage of this method is the very fast computation of the conformal
map and its inverse due to explicit formulas.

In the example presented in the main text, we set $\lambda_0 = 0.1$,
$n = 27$, and each $\theta_k$ is chosen randomly as $(2\pi/3) \eta +
0.2 \chi$, where $\chi$ is the standard normal variable (with zero
mean and unit variance), while $\eta$ takes values $0$, $1$ or $2$
with equal probabilities.  In other words, the algorithm starts from
the unit disk and then ``digs'' three long channels by progressively
removing semi-circular regions along three preferred directions $0,\,
2\pi/3,\, 4\pi/3$.

\section{Scaling of the MFPT with the area of the domain}

The MFPT averaged over uniformly distributed starting points is
known to be proportional to the area of the confining domain: $\T
\propto |\Omega|$ (see \cite{Holcman14} and references therein).
However, this scaling may not hold when the starting point is fixed.
We briefly discussed this issue in the main text by considering an
example of a thin long rectangle.  Here we extend this discussion and
explain when and why the conventional scaling may fail.

We first consider the disk of radius $R$ with an escape arc
$(-\epsilon,\epsilon)$ (see \ref{sec:disk}).  When the particles start
from the origin ($x_0 = 0$), Eq. (\ref{eq:Tdisk}) yields the MFPT
\begin{equation}
\T(0) = \frac{R^2}{D}\biggl(\frac14 - \ln \sin(\epsilon/2)\biggr) ,
\end{equation}
that indeed scales with the area of the disk.  Let now the starting
point lie near the boundary, say, at distance $a = R - |x_0|$ such
that $a \ll R$.  In this case, the Taylor expansion of
Eq. (\ref{eq:Tdisk}) in powers of $a/R$ is
\begin{equation}
\label{eq:Tdisk_Ra}
\T((R-a)e^{i\theta_0}) = \frac{1}{D}\biggl(\pi R^2 W_{\epsilon/\pi}(e^{i\theta_0}) + aR U_\epsilon(e^{i\theta_0}) + O(a^2) \biggr),
\end{equation}
where $W_{\epsilon/\pi}$ is given by Eq. (\ref{eq:Womega}), and
\begin{equation}
U_\epsilon(z) = \frac{1+z}{2\sqrt{1-2z\cos\epsilon + z^2}} \, .
\end{equation}
If the starting point lies near the escape region (i.e., $|\theta_0|
\leq \epsilon$), the function $W_{\epsilon/\pi}(e^{i\theta_0})$
vanishes, as illustrated in Fig. \ref{fig:Wtot}.  In this case, the
first term in Eq. (\ref{eq:Tdisk_Ra}), which scaled with the area $\pi
R^2$, disappears, while the next term scales {\it linearly} with the
radius $R$.  In turn, if the starting point is far from the escape
region, the MFPT scaling with the area is recovered.  We conclude that
the scaling with the area is not universal and depends on how far the
fixed starting point is from the escape region.  Since the fraction of
points near the escape region is relatively small, the average of the
MFPT over all starting points in $\Omega$ results in the conventional
scaling of the global MFPT.

The analysis for arbitrary planar domains is much more involved and
goes beyond the scope of this letter.  We just mention two possible
ways to proceed in this direction.

(i) If the original domain $\Omega$ is dilated by factor $2$ and the
original starting point $x_0$ is similarly transformed into $2x_0$,
the conformal map from the unit disk to the dilated domain is twice
the original conformal map so that one gets an additional factor $4$
from $|\phi'_{x_0}(z)|^2$ in Eq. (\ref{eq:Tfinal0}) and thus recovers
the scaling with the area.  However, this argument does not hold if
the location of the starting point in chosen differently (e.g., at a
fixed distance from the boundary).  Using the M\"obius transform, one
can move the starting point and thus investigate the scaling.

(ii) When the starting point is far from the escape region, the
harmonic measure $\omega$ is small, and the logarithmic term in
Eq. (\ref{eq:STfinalA}), which scales with the area $|\Omega|$,
provides the dominant contribution to the MFPT.  However, if the
starting point is close to the escape region, the harmonic measure is
close to $1$, and the logarithmic term vanishes.  So the main
contribution comes from the next term in Eq. (\ref{eq:STfinalA}) whose
scaling needs to be analyzed.  In analogy with the disk, a different
scaling of the MFPT can be expected in this situation.

\section{Derivation of the identity (\ref{eq:Cagi_identity})}
\label{sec:derivation}

We consider the series
\begin{equation}
\label{eq:Fseries}
Q(z,x) = 4\pi \sum\limits_{n=1}^\infty \frac{P_{n-1}(x) + P_n(x)}{2n} \, r^n \cos n\theta ,
\end{equation}
where $z = re^{i\theta}$.  Using the integral representation of
Legendre polynomials,
\begin{equation*}
P_n(x) = \frac{1}{\pi} \int\limits_0^\pi dy ~ \nu^n(y),  \quad \nu(y) = x+i\sqrt{1-x^2} \cos y, 
\end{equation*}
we have
\begin{equation*}
\begin{split}
Q & = \int\limits_0^\pi dy \biggl(1 + \frac{1}{\nu(y)}\biggr) \sum\limits_{n=1}^\infty \biggl(\frac{(r e^{i\theta} \nu(y))^n}{n} + 
\frac{(r e^{-i\theta} \nu(y))^n}{n}\biggr)  \\
& = - \int\limits_0^\pi dy \biggl(1 + \frac{1}{\nu(y)}\biggr) \biggl(\ln(1 - z \nu(y)) + 
\ln(1 - \bar{z} \nu(y))\biggr)  \\
& = -  \int\limits_0^\pi dy \biggl(1 + \frac{1}{x(1 - B\cos y)}\biggr) \biggl(\ln((1 - zx)(1 - \bar{z} x)) \\
& \qquad + \ln(1- A \cos(y)) + \ln(1 - A' \cos(y))\biggr) , \\
\end{split}
\end{equation*}
where $B = - i\sqrt{1-x^2}/x$, $A = iz \sqrt{1-x^2}/(1-zx)$, and $A' =
i\bar{z} \sqrt{1-x^2}/(1-\bar{z}x)$.  To compute the above integral,
we use the identity
\begin{equation}
\int\limits_0^\pi dy \, \frac{\ln(1 - 2a \cos y + a^2)}{1-2b\cos y + b^2} = 2\pi \, \frac{\ln(1-ab)}{1-b^2} \,,
\end{equation}
from which another identity follows
\begin{equation}
\label{eq:auxil12}
\int\limits_0^\pi dy \, \frac{\ln(1 - A\cos y)}{1-B\cos y} = \pi \,\frac{1+b^2}{1-b^2} \biggl(2\ln(1-ab) - \ln(1+a^2)\biggr) ,
\end{equation}
\break
where $A = 2a/(1+a^2)$ and $B = 2b/(1+b^2)$ or, equivalently,
\begin{equation}
\begin{split}
a &= \frac{1 - \sqrt{1-A^2}}{A} = \frac{1-zx - \sqrt{1-2xz+z^2}}{iz\sqrt{1-x^2}} \,, \\
a' &= \frac{1 - \sqrt{1-A'^2}}{A'} = \frac{1-\bar{z}x - \sqrt{1-2x\bar{z}+\bar{z}^2}}{i\bar{z}\sqrt{1-x^2}} \,, \\
b &= \frac{1 - \sqrt{1-B^2}}{B} = \frac{1-x}{i\sqrt{1-x^2}} \,.\\
\end{split}
\end{equation}

Using the relation (\ref{eq:auxil12}), we can compute all integrals
separately.  We have
\begin{widetext}
\begin{equation}
\begin{split}
-\frac{Q}{\pi} & = \ln((1 - zx)(1 - \bar{z} x)) - \ln (1+a^2) - \ln (1+a'^2) \\
& + \frac{1}{x} \biggl[\ln((1 - zx)(1 - \bar{z} x)) \frac{1+b^2}{1-b^2} + \frac{1+b^2}{1-b^2} 
\biggl(2\ln(1-ab) - \ln (1+a^2) + 2\ln(1-a'b) - \ln (1+a'^2) \biggr)\biggr]  \\
& = 2\ln((1 - zx)(1 - \bar{z} x)) + 2\ln(1-ab) + 2\ln(1-a'b) - 2\ln(1+a^2) - 2\ln(1+a'^2)  \\
& = 2\ln\biggl( \frac{(1+z-\sqrt{1-2xz+z^2})(1+\bar{z}-\sqrt{1-2x\bar{z}+\bar{z}^2})}{(1+x)^2 z\bar{z}} \frac{(1 - zx)(1 - \bar{z} x)}{(1+a^2)(1+a'^2)} \biggr)  \\
& = 2\ln\biggl(\frac{z\bar{z} (1-x)^2}{4} 
\frac{(1+z-\sqrt{1-2xz+z^2})(1+\bar{z}-\sqrt{1-2x\bar{z}+\bar{z}^2})}{(1-xz-\sqrt{1-2xz+z^2})(1-x\bar{z}-\sqrt{1-2x\bar{z}+\bar{z}^2})}  \biggr)  \\
& = 2\ln\biggl(\frac14 (1-z+\sqrt{1-2xz+z^2})(1-\bar{z}+\sqrt{1-2x\bar{z}+\bar{z}^2})  \biggr),  \\
\end{split}
\end{equation}
\end{widetext}
from which
\begin{equation}
Q(z,x) = -4\pi \ln\left(\frac{\left|1-z+\sqrt{1-2xz+z^2}\right|}{2}\right) .
\end{equation}
Comparison of this relation to Eq. (\ref{eq:Fseries}) implies the
identity (\ref{eq:Cagi_identity}).

\end{document}